\documentclass[aps,prd,twocolumn,showkeys,floats,showpacs, superscriptaddress,nofootinbib]{revtex4-1}
\usepackage{graphicx}
\usepackage{amsmath,amssymb}
\usepackage{aas_macros}
\usepackage{natbib}
\usepackage[colorlinks=true]{hyperref}

\usepackage{bm}

\def\af{\alpha_\textrm{EM}}

\begin{document}
\title{Post-Newtonian phenomenology of a massless dilaton}

\author{Aur\'elien Hees}
\email{A.Hees@ru.ac.za}
\affiliation{Department of Mathematics, Rhodes University, Grahamstown 6140, South Africa}

\author{Olivier Minazzoli}
\email{ominazzoli@gmail.com}
\affiliation{Centre Scientifique de Monaco and Laboratoire Artemis, Universit\'e C\^ote d'Azur, CNRS, Observatoire C\^ote d'Azur, BP4229, 06304, Nice Cedex 4, France}

\begin{abstract}
In this paper, we present extensively the observational consequences of massless dilaton theories at the post-Newtonian level. We extend previous work by considering a general model including a dilaton-Ricci coupling as well as a general dilaton kinetic term while using the microphysical dilaton-matter coupling model proposed in [Damour and Donoghue, PRD 2010]. 

\hspace{5mm}

We derive all the expressions needed to analyze local gravitational observations in a dilaton framework, which is useful to derive constraints on the dilaton theories. In particular, we present the equations of motion of celestial bodies (in barycentric and planetocentric reference frames), the equation of propagation of light and the evolution of proper time as measured by specific clocks. Particular care is taken in order to derive properly the observables. The resulting equations can be used to analyse a large numbers of observations: universality of free fall tests, planetary ephemerides analysis, analysis of satellites motion, Very Long Baseline Interferometry, tracking of spacecraft, gravitational redshift tests, \dots 
\end{abstract}
\pacs{04.25.Nx,04.80.Cc,04.50.-h,04.60.Cf,95.35.+d,95.36.+x}
\keywords{modified gravity, ephemerides, range, doppler, astrometry, equivalence principle, tests of gravity}

\maketitle

%-------------------------------------------------------------------------------------
\section{Introduction}

A general feature of all string and superstring theory models is the presence of a massless scalar partner to the metric called dilaton \cite{becker:2007bk,*green:1988oa,*tong:2009ar}. Perturbative calculations in the context of string theory show that the effective bosonic action should correspond to a ``general'' scalar-tensor theory \cite{callan:1986rt,damour:1994fk,*damour:1994uq}. This means that the dilaton field has gravitational strength, that it couples to the Ricci in the effective action (in the string frame) and that its kinetic term is not necessarily canonical. Besides, the dilaton is non-minimally coupled to the gauge-matter sector \cite{antoniadis:1987pl,*ferrara:1987pb,*ferrara:1987pl,*taylor:1988pb,damour:1994fk,*damour:1994uq,becker:2007bk,*green:1988oa,*tong:2009ar}. Unfortunately, the precise form of the various couplings in the effective action is not known and therefore it is impossible for now to rule out this type of theory per se. However, one can derive the rich phenomenology resulting from such non-minimal couplings and constrain them with observations and experiments. 

In \cite{damour:2008pr,damour:2010zr,*damour:2010ve}, Damour and Donoghue introduced a specific  modeling for the coupling between the dilaton and matter at the microscopic level. In addition, using new results from nuclear physics, they derived an equivalent phenomenological macroscopic modeling allowing to make theoretical predictions of equivalence principle violating effects~\cite{damour:2010zr,*damour:2010ve,damour:2008pr}. In their dilaton-matter coupling model, the dilaton couples directly to the trace-anomaly of the quantum chromodynamics (QCD) sector, which simplifies the analysis by coupling the dilaton directly to the QCD renormalization group invariant mass scale $\Lambda_3$, effectively responsible for most of the mass of hadrons \cite{gasser:1982ph}. In addition, the dilaton couples also to the fine structure constant and to the fermion masses (electron and the quarks up and down).\footnote{Effects of the mass of heavy quarks and of the weak interaction are integrated out in the chiral perturbative framework considered by Damour and Donoghue \cite{gasser:1982ph}.} In total, 5 coefficients characterize the coupling of the dilaton to matter and they can directly be related to the masses of fermions, to the QCD invariant mass scale and to the fine structure constant.

In~\cite{damour:2010zr,*damour:2010ve,damour:2008pr}, a simple gravitational part of the action formed only by the Ricci scalar (i.e. the standard Einstein-Hilbert action) is used to derive observational consequences of the dilaton-matter coupling. In particular, no Ricci-dilaton coupling or non-canonical kinetic term for the dilaton has been considered. However, actual string loop expansions seem to indicate a more complicated gravitational sector in the string frame \cite{callan:1986rt,damour:1994fk,*damour:1994uq}. Therefore, the present paper proposes to use the recent results of Damour and Donoghue regarding dilaton-matter modeling in order to derive the post-Newtonian phenomenology of a massless dilaton theory with a generalized gravitational sector. The resulting phenomenology is slightly richer and a specific situation exhibits a decoupling which leads to a strong reduction of the deviations from General Relativity (GR). This decoupling is studied separately in~\cite{minazzoli:2016rm}.

Testing GR and constraining alternative theories of gravity have been long-standing efforts in the last decades. In particular, local tests of gravitation have been widely used to test the fundamental principles of GR~\cite{schlamminger:2008zr,vessot:1979fk,*vessot:1980fk,uzan:2011vn,rosenband:2008fk,guena:2012ys,fienga:2008fk,*fienga:2009kx,*fienga:2011qf,*verma:2014jk,*fienga:2015rm,konopliv:2011dq,pitjeva:2005kx,*pitjeva:2013fk,*pitjeva:2013uq,*pitjeva:2014fj,pitjev:2013qv,hees:2012fk,iorio:2014yu,hees:2014jk,hees:2015sf}  (for a detailed review, see~\cite{will:2014la}). The main goal of this paper is to present all the relevant equations in order to analyze weak gravitational field observations within the framework of dilaton theories. Dilaton theories break the Einstein Equivalence Principle (EEP), which means that one has to be very careful when deriving observables. Deviations from GR produced by the dilaton can be classified into two classes: (i) deviations parametrized by the standard parametrized post-Newtonian (PPN) formalism~\cite{will:1993fk,will:2014la} and (ii) violations of the EEP. Both these effects need to be considered simultaneously in data analysis. In the present paper we present all the equations needed for a data analysis: equation of motion of celestial bodies (expressed either in a barycentric reference frame or in a local planetocentric reference frame), equation of light propagation and the corresponding effect on range, Doppler and astrometric observables, evolution of the proper time as measured by specific clocks. We also show how to express all the observables properly. All these results will be helpful for the implementation of data analysis in order to constrain dilaton theories correctly.

In Sec.~\ref{sec:DD}, we briefly summarized the microscopic dilaton-matter modeling from Damour and Donoghue~\cite{damour:2010zr,*damour:2010ve,damour:2008pr} and the corresponding phenomenological macroscopic model associated. In Sec.~\ref{sec:action_field}, we present the action considered in this paper. We also introduce briefly the decoupling mentioned above and fully studied in \cite{minazzoli:2016rm}. We derive the field equations and their post-Newtonian solutions. The calculations done in this section are similar to the ones that can be found in~\cite{damour:1992ys,will:1989qy,alsing:2012px}. Sec.~\ref{sec:equ_motion} is devoted to the derivation of the equations of motion of massive bodies. A particular care has been taken to rescale the gravitational constant and the masses of the bodies in order to absorb terms that are unobservable and to derive equations of motion that can be used in practice. In Sec.~\ref{sec:light}, we study the propagation of light. Both the equations of motion of massive bodies and the propagation of light are coordinates dependant. In Sec.~\ref{sec:obs}, we show how to express proper observables in dilaton theories. In particular, observables related to clocks (range and Doppler) are developed carefully since they depend on the composition of clocks used for the measurement. In Sec.~\ref{sec:use}, we present several applications where all this formalism can be used: local measurement of a differential acceleration of two test masses, test of the gravitational redshift with a comparison of two clocks, planetary ephemerides analysis, ephemerides of satellites motion (Lunar Laser Ranging, artifical satellites around Earth and around other planets and natural satellites around other planets) and Very Long Baseline Interferometry (VLBI). Sec.~\ref{sec:specific} particularizes some of the results to specific cases of the action and compares the results with the literature. Finally, we present our conclusion in Sec.~\ref{sec:conclusion}.

%-------------------------------------------------------------------------------------
\section{Damour and Donoghue dilaton-matter coupling model}\label{sec:DD}
\subsection{Linear coupling}\label{sec:DD_lin}
Let us remind the main result from \cite{damour:2010zr,*damour:2010ve}. One uses a phenomenological microscopic model for matter including a dilaton-matter coupling described by
\begin{equation}\label{eq:sm}
	S_\textrm{mat}[g_{\mu\nu},\varphi,\Psi_i]=S_\textrm{SM}[g_{\mu\nu},\Psi_i]+S_\textrm{int}[g_{\mu\nu},\varphi,\Psi_i]\, ,
\end{equation}
where $S_\textrm{SM}$ represents the Standard Model Lagrangian depending on the matter fields $\Psi_i$, $g_{\mu\nu}$ is the space-time metric and $\varphi$ is the scalar dilatonic field. The interacting part of the Lagrangian is parametrized by\footnote{Note that the scalar field $\varphi$ is dimensionless and is related to the scalar field $\phi$ used in \cite{damour:2010zr} by $\varphi=\tilde\kappa\phi$ with $\tilde\kappa^2=4\pi G/c^4$.}
\begin{eqnarray}
\label{eq:Lint}
\mathcal L_\textrm{int}=	 \varphi &\Big[&\frac{d_e}{4e^2}F_{\mu\nu}F^{\mu\nu}-\frac{d_g\beta_3}{2g_3}F^A_{\mu\nu}F_A^{\mu\nu}\\
&&-\sum_{i=e,u,d}(d_{m_i}+\gamma_{m_i}d_g)m_i\bar\psi_i\psi_i\Big] \, . \nonumber
\end{eqnarray}
with $F_{\mu\nu}$ the standard electromagnetic Faraday tensor, $e$ the electric charge of the electron, $F^A_{\mu\nu}$ the gluon strength tensor, $g_3$ is the QCD gauge coupling, $\beta_3$ denotes the $\beta$ function for the running of $g_3$, $m_i$ the mass of the fermions (electron and light quarks), $\gamma_{m_i}$ the anomalous dimension giving the energy running of the masses of the QCD coupled fermions and $\psi_i$ the fermions spinors. The coefficients $d_i$ are dimensionless constants that parametrize the interaction between the dilaton and matter. The dilaton coupling coefficients $d_i$ have a physical interpretation~\cite{damour:2010zr} since they parameterize a linear scalar field dependance of the following parameters: the fine structure constant (for $d_e$), the mass of the electron (for $d_{m_e}$), the mass of the quarks up and down (for $d_{m_u}$ and $d_{m_d}$) and the QCD mass scale $\Lambda_3$ (for $d_g$).  It has to be noted that in this effective model, the dilaton is directly coupled multiplicatively to the QCD trace anomaly $T^{\textrm{anom}}_g=\frac{\beta_3}{2g_3}\left(F^A\right)^2+\sum_{i=e,u,d}\gamma_{m_i} m_i\bar\psi_i\psi_i$ . 

Damour and Donoghue have shown that the microscopic action (\ref{eq:sm}) can be phenomenologically replaced at the macroscopic level by an action describing point masses
\begin{equation}\label{eq:smat}
		S_\textrm{mat}[g_{\mu\nu},\varphi,\Psi_i]=-c^2\sum_A \int_A d\tau~ m_A(\varphi) \, ,
\end{equation}
where $d\tau$ is the proper time defined by $c^2d\tau^2=-g_{\alpha\beta}dx^\alpha dx^\beta$. The effects produced by the coupling of the dilaton to matter is encoded in the coupling function
\begin{equation}\label{eq:alpha}
	\alpha_A(\varphi)=\frac{\partial \ln m_A(\varphi)}{\partial \varphi}\, .
\end{equation}
Within their model, they compute the expressions of the coupling coefficients $\alpha_A$ that are given by
\begin{subequations}\label{eq:alpha_DD}
 \begin{equation}\label{eq:alphaa_DD}
    \alpha_A(\varphi)\simeq d_g^* + \bar \alpha_A \, ,
 \end{equation}
 where $d_g^*$ is a constant independant of the composition of the body given by
 \begin{equation}\label{eq:dg_s}
  d_g^*\simeq d_g+0.093 (d_{\hat m}-d_g)+0.000\ 27\ d_e\, ,
 \end{equation}
and where $\bar \alpha_A$ depends on the constants $d_i$ and on the physical properties (or composition) of the body $A$. The semi-analytical expression of $\bar\alpha_A$ has been fully derived in~\cite{damour:2010zr,*damour:2010ve} and reads
\begin{equation}
\bar{\alpha}_A \simeq  \left[(d_{\hat{m}} - d_g) Q'_{\hat{m}} + d_e Q'_e \right]_A \, , \label{eq:alphabarsemianal}
\end{equation}
where $d_{\hat{m}}$ corresponds to the contribution of the symmetric combination of the light-quark masses
\begin{equation}
d_{\hat{m}}= \frac{d_{m_d} m_d+d_{m_u} m_u}{m_d+ m_u}
\end{equation}
and where the ``dilatonic charges'' read
\end{subequations}
\begin{subequations}\label{eq:dil_charges}
\begin{equation}
Q'_{\hat{m}}= - \frac{0.036}{A^{1/3}}- 1.4 \times 10^{-4} \frac{Z(Z-1)}{A^{4/3}},
\end{equation}
and
\begin{equation}
Q'_e = + 7.7 \times 10^{-4} \frac{Z(Z-1)}{A^{4/3}}, \label{eq:chargee}
\end{equation}
\end{subequations}
where A and Z are the mass and atomic numbers respectively. It has to be stressed that the approximation (\ref{eq:alpha_DD}) is not suitable for single nucleons or light  isotopes. For them, one should rather use the complete formula given in \cite{damour:2010zr,*damour:2010ve} (see Appendix \ref{app:alphacoeff}). 

This result is general and independent of the gravitational part of the action. The important point lies in the correspondance between the microphysics action~(\ref{eq:sm}) suitable to describe microscopic interaction and the macrophyiscs phenomenological action~(\ref{eq:smat}) suitable for deriving equations of motion and osbervables. However, one has to notice that the action (\ref{eq:sm}) is already effective (or phenomenological) at some level since the QCD trace anomaly appears explicitely in it.

\subsection{Non-linear coupling} \label{sec:DD_nonlin}
A straightforward generalization of the Damour-Donoghue model is to consider non-linear couplings between the dilaton and matter. The corresponding interacting part of the Lagrangian would therefore write
\begin{eqnarray}\label{eq:interLagrangeNL}
 \mathcal L_\textrm{int}&=&	  \Bigg[\frac{D_e(\varphi)}{4e^2}F^2-\frac{D_g(\varphi)\beta_3}{2g_3}\left(F^A\right)^2\\
&& \quad -\sum_{i=e,u,d}\Big(D_{m_i}(\varphi)+\gamma_{m_i}D_g(\varphi)\Big)m_i\bar\psi_i\psi_i\Bigg] \, , \nonumber
\end{eqnarray}
where $D_i(\varphi)$ are functions of the scalar field. We suppose here that $D_i(\varphi_0)=0$ with $\varphi_0$  the background value of the scalar field. A non-vanishing value of $D_i(\varphi_0)$ would simply lead to a rescaling   of the five parameters: fine structure constant, masses of the electron and of the up and down quarks and of the QCD mass scale $\Lambda_3$. 

In this case, one can replace the matter action by a standard point particle whose mass depends on the scalar field $m_A(\varphi)$ through the quantity
\begin{equation}
 \alpha_A(\varphi)=\frac{\partial \ln m_A(\varphi)}{\partial \varphi}  \, .
\end{equation}
The results of Damour and Donoghe now becomes
\begin{subequations}\label{eq:alpha_NL}
\begin{equation}
 \alpha_A(\varphi)\simeq {D_g^*}'(\varphi) + \bar \alpha_A(\varphi)  \, ,
\end{equation}
where the prime denotes the derivative with respect to $\varphi$, 
\begin{equation}\label{eq:DG*}
 {D_g^*}'(\varphi)=D_g'(\varphi)+0.093\big(D_{\hat{m}}'(\varphi)-D_g'(\varphi)\big)+0.000 \ 27 D_e'(\varphi) \, ,
\end{equation}
and 
\begin{equation}\label{eq:bar_alpha_A}
\bar\alpha_A(\varphi)= \Big[\big(D_{\hat{m}}'(\varphi)-D_g'(\varphi)\big)Q'_{\hat m}+D_e'(\varphi)Q'_e\Big]_A\, ,
\end{equation}
with the dilatonic charges defined by Eqs.~(\ref{eq:dil_charges}).

The post-Newtonian phenomenology of a dilaton non-linearly coupled to matter will depend on a new quantity (that vanishes in the case of linear coupling) given by
\begin{equation}
 \beta_A(\varphi)=\frac{\partial \alpha_A(\varphi)}{\partial \varphi}\simeq {D_g^*}''(\varphi) + \bar \beta_A(\varphi)  \, ,
\end{equation}
with 
\begin{equation}\label{eq:bar_beta}
 \bar\beta_A(\varphi)=\Big[\big(D_{\hat{m}}''(\varphi)-D_g''(\varphi)\big)Q'_{\hat m}+D_e''(\varphi)Q'_e\Big]_A\, .
\end{equation}

\end{subequations}

%In the following, for deriving the post-Newtonian phenomenology of this kind of model, we will need to evaluate the previous quantities at the background scalar field value $\varphi_0$. Therefore, in the following, we will use the notations
%\begin{subequations}
 %\begin{align}
 % d_i&=D'_i(\varphi_0) \, ,\\
 % d'_i&=D''_i(\varphi_0) \, .
 %\end{align}
 %Therefore $\bar\alpha(\varphi_0)=\bar\alpha_A$ is given by Eq.~(\ref{eq:alphabarsemianal}) and
 %\begin{align}
 % {d_g^*}'&={D_g^*}''(\varphi_0)=d_g'+0.093(d_{\hat{m}}'-d_g')+0.000\ 27 d_e' \, ,\\
 % \bar\beta_A&=\bar\beta_A(\varphi_0)=\left[ (d_{\hat{m}}'-d_g')Q'_{\hat{m}} + d'_e Q'_e\right]_A \, .
 %\end{align}
%\end{subequations}

%-------------------------------------------------------------------------------------
\section{Action, field equations and their post-Newtonian solutions}\label{sec:action_field}

In the following, we consider a general gravitational part in the action, such that the action reads
\begin{eqnarray}
	S&=&\frac{1}{c}\int d^4x \frac{\sqrt{-g}}{2\kappa}\left[f(\varphi)R-\frac{\omega(\varphi)}{\varphi}g^{\mu\nu}\partial_\mu\varphi\partial_\nu\varphi\right] \nonumber\\
	&& \qquad + S_\textrm{mat} [g_{\mu\nu},\varphi,\Psi_i]  \, , \label{eq:actionstringframe}
	%&&+ \frac{1}{c}\int d^4x \sqrt{-g}\mathcal L_\textrm{mat}[g_{\mu\nu},\varphi,\Psi_i]  \, , \label{eq:actionstringframe}
\end{eqnarray}
with $\kappa=8\pi G/c^4$, $R$ the Ricci scalar and $f(\varphi)$ and $\omega(\varphi)$ two functions of the scalar field. The gravitational part of this action is a generalization of the one used by Damour and Donoghue \cite{damour:2010zr,*damour:2010ve}, in which $f(\varphi)=1$ and $\omega(\varphi)=2\varphi$. Except for the scalar-matter coupling, it corresponds to a general Brans-Dicke action \cite{brans:1961fk,*brans:2014sc}. The matter part of the action is given by Eq.~(\ref{eq:sm}) with the matter-dilaton interaction described by the generalization of Damour-Donoghue model~\cite{damour:2010zr} parametrized by Eq.~(\ref{eq:interLagrangeNL}). As mentioned in Sec.~\ref{sec:DD}, the matter action can be replaced phenomenologically by the action~(\ref{eq:smat}) with Eq.~(\ref{eq:alpha_NL}). One can derive and solve the field equations in this frame,\footnote{The word ``frame'' is sometimes replaced by the word ``representation'' in the literature in order to mark the difference between conformal frames and reference frames. In this paper, the word ``frame'' alone refers to a conformal frame. A reference frame will be explicitely named so in order to avoid any confusion.} which is sometimes known as the ``string frame''. Instead, it is more convenient to perform a conformal transformation and solve the field equations in the so-called Einstein frame where calculations are much easier~\cite{damour:1992ys}.

\subsection{On a specific model with a decoupling}\label{sec:decoupling}
Sec.~\ref{sec:DD} suggests to introduce a new decomposition of the phenomenological mass $m_A(\varphi)$ of particles as
\begin{equation}\label{eq:dec_mass}
 m_A(\varphi)=e^{D_g^*(\varphi)}\bar m_A(\varphi) \, ,
\end{equation}
where the first term is identical for all bodies and is composition independant and the second part is entirely composition dependant. This decomposition is coherent with Eqs.~(\ref{eq:alphaa_DD}) and (\ref{eq:alpha_NL}) with
\begin{equation}\label{eq:bar_alpha_a_m}
 \frac{\partial\ln \bar m_A(\varphi)}{\partial \varphi}= \bar\alpha_A(\varphi) \, .
\end{equation}
 One can introduce a new metric conformally related to the original metric $g_{\mu\nu}$ through
\begin{equation}
 \bar g_{\mu\nu} = e^{2D_g^*(\varphi)} g_{\mu\nu} \, .
\end{equation}
In this frame, the action becomes
\begin{align}
 S&=\frac{1}{c}\int d^4x \frac{\sqrt{-\bar g}}{2\kappa}\left[e^{-2D_g^*(\varphi)}f(\varphi)\bar R-\right. \nonumber \\
 &\qquad \left.e^{-2D_g^*(\varphi)}~ \left(6\big({D_g^*}'(\varphi)\big)^2+\frac{\omega(\varphi)}{\varphi}\right)\bar g^{\alpha\beta}\partial_\alpha\varphi\partial_\beta\varphi\right] \nonumber \\
 &\qquad -c^2 \sum_A \int_A d\, \bar\tau \, \bar m_A(\varphi) \, . \label{eq:action_decoupled}
\end{align}
This frame corresponds to a frame in which matter is the least coupled (to the scalar field), such that the coupling is effective only through terms that are composition dependent. 

From this observation, it is important to realize that there is a specific model where the original Ricci-scalar coupling $f(\varphi)$ can be completely absorbed by the composition independent matter-scalar coupling. Indeed, this specific model is characterized by
\begin{equation}\label{eq:decoupling}
 f(\varphi)=e^{2D_g^*(\varphi)} \, .
\end{equation}
The consequences of this equality are developed thoroughly in~\cite{minazzoli:2016rm}. But one should note that in this case, the newly introduced frame corresponds exactly with the Einstein frame (see Sec.~\ref{sec:einstein_frame}). 

Note that an expansion of the condition (\ref{eq:decoupling}) around the scalar field background value $\varphi_0$ gives
\begin{subequations}
 \begin{align}
  f_0 &= e^{2 D_{g0}^*} \, ,\\
  \frac{f'_0}{f_0}&= 2 {D_{g0}^*}' \, ,\\
  \frac{f''_0}{f_0}-\left(\frac{f'_0}{f_0}\right)^2&=2 {D_{g0}^*}'' \, ,
 \end{align}
where the subscripts 0 will refer to quantities depending on the scalar field evaluated at its background value: $X_0=X(\varphi=\varphi_0)$.
\end{subequations}
We shall see that these conditions lead to a decoupling in the post-Newtonian dilaton phenomenology which therefore tends to reduce strongly deviations from GR~\cite{minazzoli:2013fk,minazzoli:2015ax} (see Sec.~\ref{sec:Phendecoupling}).

 Finally, it is important to realize that there is a rescaling of all units related to the conformal transformation. In particular, the background value of the gravitational constant in this frame units will be rescaled as~\cite{dicke:1962ly}
\begin{equation}\label{eq:gbar}
 \bar G = e^{2D_g^*(\varphi_0)}G\, .
\end{equation}

 %-------------------------------------------------------------------------------------
\subsection{Einstein frame action and field equations}\label{sec:einstein_frame}

One can use a conformal transformation in order to write the action in a form where the kinetic part related to the metric is the one of General Relativity. Such a conformal frame is called the Einstein frame. Moreover, one can rescale the scalar-field such that its kinetic part becomes canonical. The action in this frame with a rescaled scalar field $\phi$ reads
\begin{eqnarray}
 S&=&\frac{1}{c}\int d^4x \frac{\sqrt{-g_*}}{2\kappa_*}\left[R_* - 2g_*^{\mu\nu}\partial_\mu \phi\partial_\nu \phi\right] \nonumber \\
 &+& \frac{1}{c}\int d^4x \sqrt{-g_*}\mathcal L^*_\textrm{mat}\left[\frac{f_0}{f(\varphi)}g^*_{\mu\nu},\phi,\Psi_i \right]
\end{eqnarray}
where the stars denote quantities in the Einstein frame and where the conformal transformation is defined by 
\begin{subequations}
 \begin{align}
    g_{\mu\nu}^* &= \frac{f(\varphi)}{f_0} g_{\mu\nu} \, ,\\
    \mathcal L^*_\textrm{mat}&= \left(\frac{f_0}{f(\varphi)}\right)^2 \mathcal L_\textrm{mat} \, ,\\
    \kappa_*=\frac{8\pi G_*}{c^4}&=\frac{\kappa}{f_0}=\frac{8\pi G}{c^4 f_0} \, ,
    \end{align}
\end{subequations}
and where we have introduced a rescaled scalar field $\phi$ which depends on the original one by
\begin{equation}\label{eq:rescaled_sf}
 \frac{d\phi}{d\varphi}=\sqrt{\frac{Z(\varphi)}{2}}\, ,
\end{equation}
with 
\begin{equation}
 Z(\varphi)=\frac{\omega(\varphi)}{\varphi f(\varphi)}+\frac{3}{2}\left(\frac{f'(\varphi)}{f(\varphi)}\right)^2 \, . \label{eq:defZ}
\end{equation}
The resulting field equations write~\citep{damour:1992ys,damour:1994fk}
\begin{subequations}\label{eq:field_eq}
\begin{eqnarray}
  R^*_{\mu\nu} &=&\kappa_* \left(T_{\mu\nu}^*-\frac{1}{2}g^*_{\mu\nu} T^*\right) + 2 \partial_\nu \phi \partial_\mu\phi \, , \label{eq:metriceqEinstein} \\
  \Box_* \phi &=& -\frac{\kappa_*}{2}\sigma^* \,
 \end{eqnarray}
\end{subequations} 
 where
 \begin{subequations}\label{eq:stress_energy}
 \begin{align}
 T^*_{\mu\nu}&=-\frac{2}{\sqrt{-g_*}}\frac{\delta \sqrt{-g_*}\mathcal L^*_\textrm{mat}}{\delta g^{\mu\nu}_*}\, ,\\
\sigma^*&=\frac{\delta \mathcal L^*_\textrm{mat}}{\delta \phi}\, .
\end{align}
\end{subequations}

 %-------------------------------------------------------------------------------------
\subsection{Bodies as point particles}

As mentioned in Sec.~\ref{sec:DD}, one can use a point particle action at the macroscopic level which can be written as
\begin{equation}
S_\textrm{mat} = - c^2 \sum_A \int_A m_A (\varphi) d\tau = - c^2 \sum_A \int_A m^*_A (\varphi) d\tau_* \, , \label{eq:matactionSF}
\end{equation}
where the two masses are related to each others by
\begin{equation}\label{eq:masses_frames}
m^*_A(\varphi)= \sqrt{\frac{f_0}{f(\varphi)}}~ m_A(\varphi)\, .
\end{equation}
Note that in particular $m_A^*(\varphi_0)=m_A(\varphi_0)$. This matter action can be written from a Lagrangian density
\begin{equation}
 \mathcal L_\textrm{mat}^*=- \sum_A \rho^*_A=-\sum_A \frac{c^2m^*_A(\varphi(x^\mu))}{\sqrt{-g_*}u^0_{*}}\delta^{(3)}(x^i-x^i_A(x^0)) \, ,
\end{equation}
where $\rho^*_A$ is the energy density,  $u_*^\alpha= dx^\alpha/ cd\tau_*$ is the Einstein frame 4-velocity, $x^i_A$ is the trajectory of the particle $A$ while $\delta^{(3)}$ is the 3D Dirac distribution.

This matter modelling leads to the following expressions of the source terms~(\ref{eq:stress_energy}) of the field equations~(\ref{eq:field_eq})
\begin{subequations}
	\begin{eqnarray}
		T_*^{\mu\nu}&=&\sum_AT^{\mu\nu}_{*A}=\sum_A \rho^*_A u^\mu_{*A} u^\nu_{*A} \, ,\\
		T_* &=&\sum T^*_A =-\sum \rho^*_A =\mathcal L_\textrm{mat}^* \, ,\\
		\sigma^*&=&-\sum_A \rho^*_A \alpha^*_A=\sum_A \alpha^*_A T^*_A \, ,
	\end{eqnarray}
\end{subequations}
where
\begin{equation}
 \alpha^*_A(\phi)=\frac{\partial \ln m_A^*(\phi)}{\partial \phi}=\frac{\partial \ln m_A^*(\phi)}{\partial \varphi}\sqrt{\frac{2}{Z(\varphi)}} \, .
\end{equation}
Using the relation (\ref{eq:masses_frames}) and the decomposition from Eqs.~(\ref{eq:dec_mass}) and (\ref{eq:bar_alpha_a_m}), one can write
\begin{subequations}\label{eq:alpha_s}
 \begin{equation}
 \alpha^*_A(\phi)=\alpha(\phi) + \tilde \alpha_A(\phi) \, ,
\end{equation}
with
\begin{align}
 \alpha(\phi)&= \sqrt{\frac{2}{Z(\varphi)}}\left[ {D_g^*}'(\varphi)-\frac{1}{2}\frac{f'(\varphi)}{f(\varphi)}\right]\, \label{eq:alpha_phi}\\
 \tilde\alpha_A(\phi)&=\sqrt{\frac{2}{Z(\varphi)}} \bar\alpha_A(\varphi) \, \label{eq:tilde_alpha},
\end{align}
\end{subequations}
where $\varphi$ now depends on $\phi$ through Eq.~(\ref{eq:rescaled_sf}).  One can directly see that the decoupling mentioned in Sec.~\ref{sec:decoupling} leads to $\alpha(\phi)=0$. 

The term $\alpha(\phi)$ is coming from the Ricci-scalar coupling and from the composition independent part of the matter-scalar coupling. The second part  $\tilde\alpha_A(\phi)$ is coming from the composition dependent part of the matter-coupling and is responsible for the violation of the EEP. Obviously, if this last term vanishes the theory will not exhibit any violation of the universality of free fall.

 %-------------------------------------------------------------------------------------
\subsection{Post-newtonian solution to the field equations}
One of the main advantages of the Einstein frame lies in the simplicity of calculations to solve the field equations. In particular, using a similar method as the one used in~\cite{damour:1991tp,*damour:1992lp,*damour:1993kc,*damour:1994fd,damour:1991tb,damour:1992ys,brumberg:1989fk,kopeikin:2004qb,soffel:2003bd,minazzoli:2009fk,minazzoli:2011fk}, one can show that the post-Newtonian metric can be parametrized as follows
\begin{subequations}\label{eq:metricEfT}
	\begin{eqnarray}
		g^*_{00}&=&-1 +2 \frac{w_*}{c^2} -2 \frac{w^2_*}{c^4}  + \mathcal O(1/c^6) \, \label{eq:metricEf00}\, ,\\
		g^*_{0i}&=&-4\frac{w^i_*}{c^3}  +\mathcal O(1/c^5) \, ,\\
		g^*_{ij}&=&\delta_{ij}\left(1 +2  \frac{w_*}{c^2}\right)+\mathcal O(1/c^4) \, \label{eq:metricEfij} \, .
	\end{eqnarray}
\end{subequations}
The post-Newtonian gravitational potential $w_*$ and $w_*^i$ are then solution of the field equations and, using the harmonic gauge,\footnote{One should note that the harmonic gauge is imposed here in the Einstein frame while other authors may have been imposing it in other frames (see \cite{klioner:2000sh} for instance). It has to be noted that our choice is based on the simplicity of the field equations in the former case. Indeed, imposing the harmonic gauge in any other frame than the Einstein one would be incompatible with the simple form of the metric (\ref{eq:metricEfT}) which satisfies the strong isotropy condition defined in \cite{damour:1991tp,*damour:1992lp,*damour:1993kc,*damour:1994fd}, and the field equations would therefore be appreciably more complicated. See for instance the discussion in \cite{minazzoli:2011fk}.} are given by (the derivation can be found for instance in~\cite{damour:1992ys})
\begin{widetext}
\begin{subequations}\label{eq:metric_einstein}
	\begin{eqnarray}
		w_*( x^\mu)&=& w_{0*} -\frac{1}{c^2}\Delta_* + \mathcal O(1/c^4)\, ,\\
		w_*^j(x^\mu)&=&\sum_A \frac{G_*m^*_{A0}v^j_A}{r_A}+\mathcal O(1/c^2)  \, ,	\\
		\phi&=&\phi_0 - \sum_A \alpha^*_{A0}\frac{G_* m_{A0}^*}{c^2 r_A } + \sum_A \frac{G_*m_{A0}^*}{2c^4  r_A}\alpha^*_{A0}\left[  \bm r_A.\bm a_A+\frac{(\bm r_A.\bm v_A)^2}{r_A^2} \right] \nonumber  \\
&&\qquad + \sum_A\frac{G_*m_{A0}^*}{r_Ac^2} \sum_{B\neq A}\frac{G_*m^*_{B0}}{r_{AB}c^2} \left[ \alpha^*_{A0}+{\alpha^*}^2_{A0}\alpha_{B0}^*+\beta_{A0}^*\alpha_{B0}^*\right]+\mathcal O(1/c^6)\, . \label{eq:scalar_EF}%\\
	\end{eqnarray}
\end{subequations}
\end{widetext}
where $w_{0*}$ is the Newtonian potential
\begin{equation}
 w_{0*}=\sum_A \frac{G_* m_{A0}^*}{r_A} \, ,
\end{equation}
and  $\Delta^*$ is a generalization of the $\Delta$ appearing in the IAU conventions~\cite{soffel:2003bd}
\begin{eqnarray}
 \Delta^*&=&\sum_A \frac{G_*m_{A0}^*}{r_A}\left[\frac{\bm r_A.\bm a_A}{2}+\frac{(\bm r_A.\bm v_A)^2}{2r_A^2}-2v_A^2\right. \nonumber\\
 &&\qquad \left.+\sum_{B\neq A}\left(1+\alpha^*_{A0}\alpha^*_{B0}\right)\frac{G_*m_{B0}^*}{r_{AB}}\right]\, .
\end{eqnarray}

In the previous expressions, $m^*_{A0}=m^*_A(\varphi_0)=m_{A0}$, $\bm r_A=\bm x-\bm x_A(x^0)$, $r_A=\left|\bm r_A\right|$, $r_{AB}=\bm x_B-\bm x_A$, $\bm v_A= d\bm x_A/dt$ and $\bm a_A=d\bm v_A/dt$. 

Finally, the $\beta^*_A$ coefficient that appears in the expression of the scalar field is given by
\begin{subequations}\label{eq:beta_s}
\begin{eqnarray}
 \beta^*_A(\phi)=\frac{\partial \alpha^*_A(\phi)}{\partial \phi}&=&\beta(\phi)+\tilde \beta_A(\phi) \, ,
\end{eqnarray}
with
\begin{align}
 \beta(\phi)=&\frac{\partial \alpha(\phi)}{\partial \phi}\nonumber \\
 =&\frac{2}{Z(\varphi)}\left[{D_g^*}''(\varphi)-\frac{1}{2}\frac{f''(\varphi)}{f(\varphi)}+\frac{1}{2}\left(\frac{f'(\varphi)}{f(\varphi)}\right)^2\right] \, \label{eq:beta_phi}\\
 &\qquad-\frac{Z'(\varphi)}{\sqrt{2} Z^{3/2}(\varphi)}\alpha(\phi) \, ,\nonumber\\
 \tilde \beta_A(\phi)=&\frac{\partial\tilde \alpha_A(\phi)}{\partial \phi} \nonumber\\
 =&\frac{2}{Z(\varphi)}\bar\beta_A(\varphi) - \frac{Z'(\varphi)}{\sqrt{2}Z^{3/2}(\varphi)} \tilde\alpha_A(\phi) \, ,\label{eq:tilde_beta}
\end{align}
\end{subequations}
with the $\alpha$ and $\tilde\alpha_A$ coefficients defined by Eqs.~(\ref{eq:alpha_s}) and $\bar\beta_{A}$ defined by Eq.~(\ref{eq:bar_beta}). 

\section{Equations of motion}\label{sec:equ_motion}

%-------------------------------------------------------------------------------------
\subsection{Einstein-Infeld-Hoffmann Lagrangian and equations of motion}\label{sec:eom}
The equations of motion of a test mass $T$ can be derived using a variational principle from the matter action~(\ref{eq:matactionSF}). This procedure is similar to the one used in~\cite{will:1989qy,damour:1992ys,will:1993fk,alsing:2012px}. The corresponding Lagrangian is given by
\begin{equation}\label{eq:lag_start}
 L_T=-m^*_T(\phi)\left(-g_{00}^*-2g_{0i}^*\frac{v^i_T}{c}-g_{ij}^*\frac{v_T^iv_T^j}{c^2}\right)^{1/2} \, .
\end{equation}
The post-Newtonian expression of this Lagrangian is obtained by expanding $m^*_T(\phi)$ and using the post-Newtonian solutions of the scalar field and the space-time metric given by Eqs.~(\ref{eq:metric_einstein}). The Newtonian part is therefore given by 
\begin{equation}\label{eq:Lag_New}
 L_{T,\textrm{N}}=m_{T0}\frac{v_T^2}{2}+m_{T0} W_T\, ,
\end{equation}
where we introduced a novel Newtonian potential
\begin{equation}\label{eq:WT}
	W_T=\sum_{A \neq T} \frac{G_{AT}m_{A0}}{r_{AT}} 
\end{equation}
that depends on the following gravitational constant (see also~\cite{will:1989qy,damour:1992ys})
\begin{equation}\label{eq:gat}
 G_{AT}=G_* \left(1+\alpha_{A0}^*\alpha_{T0}^*\right)=\frac{G}{f_0}\left(1+\alpha_{A0}^*\alpha_{T0}^*\right) \, .
\end{equation}
The post-Newtonian contribution to the Lagrangian is given by
\begin{equation}\label{eq:pnLagrangian}
  \frac{c^2}{m_{T0}} L_{T,\textrm{pN}}=\frac{v^4}{8} -4 v^i_T W^i_T +\frac{v^2}{2} W_T + v^2 U_T -\frac{W_T^2}{2}-\Delta_T \, ,
\end{equation}
where we have now introduced
\begin{subequations}
 \begin{align}
  U_T &= \sum_{A \neq T}\tilde \gamma_{AT} \frac{G_{AT} m_{A0}}{r_{AT}} \, , \\
  W^i_T&=\sum_{A \neq T}\frac{1+ \tilde\gamma_{AT}}{2} \frac{G_{AT} m_{A0}v^i_A}{r_{AT}} \, , \\
\Delta_T&= \,\sum_{A \neq T}\frac{G_{AT}m_{A0}}{r_{AT}}\left[-(\tilde\gamma_{AT}+1)v_A^2+\frac{1}{2}\bm r_{AT}.\bm a_A\right. \nonumber\\
& \left. +\frac{(\bm v_A.\bm a_A)^2}{2r_{AT}^2}+\sum_{B\neq A}(2 \tilde\beta^A_{BT}-1)\frac{G_{AB}m_{B0}}{r_{AB}}\right] \\
&+\sum_{A\neq T,B \neq T} (\tilde\beta^T_{AB}-1)\frac{ G_{AT}m_{A0}}{r_{AT}}\, \frac{G_{BT}m_{B0}}{r_{BT}} \, , \nonumber\\
  \tilde\gamma_{AT}&= \frac{1-\alpha^*_{A0}\alpha^*_{T0}}{1+\alpha^*_{A0}\alpha^*_{T0}} \, , \label{eq:gamma_AT}\\
\tilde \beta^T_{AB}-1&=\frac{\beta_{T0}^*}{2}\frac{\alpha_{A0}^*}{1+\alpha_{A0}^*\alpha_{T0}^*}\frac{\alpha_{B0}^*}{1+\alpha_{B0}^*\alpha_{T0}^*} \, ,
 \end{align}
\end{subequations}
where the coefficients $\alpha^*$ are defined by Eqs.~(\ref{eq:alpha_s}) and $\beta^*$ are defined by Eq.~(\ref{eq:beta_s}).
 
%\subsection{Equations of motion of particles}
%\label{sec:eqm}
From the previous Lagrangian, the equations of motion for a test particle $T$ can easily be derived. They read
%\begin{widetext}
\begin{align}
\bm {a}_T&=\bm{\nabla} W_T+\frac{1}{c^2}\Bigg[  v_T^2 \bm{\nabla} U_T-2 (W_T+U_T) \bm \nabla  W_T-\bm \nabla \Delta_T \nonumber\\
&-2   \left(\bm{v}_T.\bm{\nabla} W_T\right)\bm{v}_T-2   \left(\bm{v}_T.\bm{\nabla} U_T\right)\bm{v}_T +4\partial_t  \bm {  W_T} \label{eq:eqm} \\
&+4\left(\bm{\nabla}\times\bm{ W_T}\right)\times\bm{v}_T-\bm v_T\partial_t W_T -2\bm v_T\partial_t U_T  \Bigg] \, . \nonumber
\end{align}
%\end{widetext} 
This equation is a generalization of the EIH (Einstein-Infeld-Hoffmann) equation of motion. 

 %-------------------------------------------------------------------------------------
\subsection{Simplifications for practical use}
\label{sec:simplifications}
For most of applications (like tracking of spacecrafts or planetary ephemeride analysis for example), the equations of motion presented previously can be simplified. First of all, we will use the decomposition of the coefficients $\alpha^*_{A0}$ and $\beta^*_{A0}$ introduced in Eqs.~(\ref{eq:alpha_s}) and (\ref{eq:beta_s}) into two parts: a part that is composition independent $\alpha_0$ defined by Eq.~(\ref{eq:alpha_phi}) and $\beta_0$ defined by Eq.~(\ref{eq:beta_phi})  (coming from the conformal transformation but also from the composition independent part of the dilaton-matter coupling) and a part that is composition dependent $\tilde \alpha_{A0}$ from Eq.~(\ref{eq:tilde_alpha}) and $\tilde \beta_{A0}$ from Eq.~(\ref{eq:tilde_beta}).
In the case of the linear Damour-Donoghue coupling presented in Sec.~\ref{sec:DD_lin}, the coefficient $\tilde \beta_{A0}$ vanishes. It is also interesting to remark that the decoupling introduced in Sec.~\ref{sec:decoupling} leads to vanishing coefficients $\alpha_0$ and $\beta_0$.

\subsubsection{Simplification at the Newtonian level}
This decomposition is important in order to identify the observed gravitational parameters $\tilde G \tilde m_{A}$ of the different bodies. These gravitational parameters are determined through the motion of bodies around other bodies (for example through the motion of planets around the Sun or through the motion of satellites around a planet). Therefore, these gravitational parameters need to be identified in the Newtonian potential, i.e. in $W_T$ given by Eq.~(\ref{eq:WT}). Let us decompose the gravitational constant $\tilde G_{AT}$ from Eq.~(\ref{eq:gat}) as follows
\begin{equation}\label{eq:gat_2}
 G_{AT} = \tilde G\left(1+\delta_A+\delta_T+\delta_{AT}\right) \, ,
\end{equation}
where we introduced
\begin{subequations}\label{eq:deltas}
 \begin{align}
  \tilde G &= G_*\left(1+\alpha_0^2\right)=\frac{G}{f_0}\left(1+\alpha_0^2\right) \, , \label{eq:tildeG}\\
  \delta_A &= \frac{\alpha_0\tilde\alpha_{A0}}{1 + \alpha_0^2 } \, ,\label{eq:delta_A}\\
  \delta_{AT} &= \frac{\tilde\alpha_{A0}\tilde\alpha_{T0}}{1 + \alpha_0^2 }\label{eq:delta_AT} \, .
 \end{align}
\end{subequations}

Note that at the Newtonian level, the equations of motion can be derived from the Lagrangian
\begin{equation}
 L_{T,\textrm{N}} =  m_{T0} \frac{v_T^2}{2} + \sum_{A\neq T} \frac{\tilde G m_{A0} m_{T0}}{2r_{AT}}\left(1+\delta_A+\delta_T+\delta_{AT}\right) \, .\label{eq:EIHLN}
\end{equation}

Usually, a violation of the universality of free fall is parametrized in terms of a difference between the inertial mass $m^I$ and  the gravitational mass $m^G$ that reads as follows \cite{will:1993fk,will:2014la}
\begin{equation}\label{eq:mg_mi}
 \frac{m^G_A}{m^I_A}=1+\delta_A \, .
\end{equation}
But it is important to notice that Eq. (\ref{eq:EIHLN}) shows that this usual parametrization is not appropriate in general for dilatonic theories because of the $\delta_{AB}$ coefficients.\footnote{Indeed, one should be careful with this expression when $\alpha_0=0$, because then $\delta_A$ vanishes while $\delta_{AB}$ may not be zero.} 

In general, one can write
\begin{eqnarray} 
\delta_{AB}&=&\delta_A \frac{\tilde \alpha_{B0}}{\alpha_0}=\delta_A ~\frac{\bar{\alpha}_{B0}}{{D_{g0}^*}' - f'_0/2f_0 }\, .\label{eq:delta_AB_comp} 
\end{eqnarray} 
It is interesting to mention that ${D_{g0}^*}' \gg \bar\alpha_{B0}$ because of the  coefficients involved in the dilatonic charges~(\ref{eq:dil_charges}) that appears in the expression of $\bar\alpha_{B0}$ (\ref{eq:bar_alpha_A}) are small compared to the ones in Eq.~(\ref{eq:DG*}). This reflects the fact that most of the baryonic mass comes from the gluonic confinment energy that is common to all baryons~\cite{damour:2008pr,damour:2010zr,*damour:2010ve}. Eq.~(\ref{eq:delta_AB_comp}) shows that in most cases (if there is no cancellation in the denominator), this leads to  $\delta_{AB} \ll \delta_A$. Nevertheless, the decoupling developed in Sec.~\ref{sec:decoupling} leads to $\alpha_0=0$. In that particular case, $\delta_A=\delta_T=0$, while $\delta_{AT}\neq 0$ cannot be neglected and has to be considered \footnote{In the decoupling limit, $\delta_{AT}\neq 0$ is not obvious from Eqs. (\ref{eq:delta_AB_comp}) because in this form, $\delta_{AB}$ is written as a quotient of two terms that go to zero at the decoupling limit. One should rather look at Eq. (\ref{eq:delta_AT}).}.

In addition to the rescaling of the gravitational constant introduced in Eq.~(\ref{eq:tildeG}), the following rescaling of the masses
\begin{equation}\label{eq:tilde_m}
	\tilde m_A=m_{A0}(1+\delta_A),
\end{equation}
is unobservable. Using this rescaling leads to the Newtonian Lagrangian 
\begin{equation}
 L_{T,\textrm{N}}=  \tilde m_{T}\left(1-\delta_T\right) \frac{v_T^2}{2} +\tilde m_{T} W_T \, ,
\end{equation}
where the modified Newtonian potential is now given by
\begin{equation} \label{eq:modeNewton}
	\tilde W_T= \sum_{A \neq T} \frac{\tilde G \tilde m_{A}}{r_{AT}}\left(1+\delta_{AT}\right)\, .
\end{equation}
Masses $\tilde m_{A}$ can be identified to what is sometimes called ``(passive) gravitational masses''.

%-------------------------------------------------------------------------------------
\subsubsection{Simplification at the post-Newtonian level}\label{sec:eom_pn_s}
We can introduce the notations from Eqs.~(\ref{eq:gat_2}), (\ref{eq:deltas}) and~(\ref{eq:tilde_m}) into the first post-Newtonian contribution to the Lagrangian from Eq.~(\ref{eq:pnLagrangian}). Since the coefficients $\delta_A$ and $\delta_{AT}$ are already appearing at the Newtonian level, we can safely neglect them at the first post-Newtonian order. This considerably simplify the post-Newtonian Lagrangian that can be written as
\begin{align}\label{eq:pnLagrangian_simp}
  \frac{c^2}{\tilde m_T} L_{T,\textrm{pN}}&=\frac{v_T^4}{8} -2 (1+\tilde\gamma)v^i_T W^i +\frac{(1+2\tilde\gamma)}{2}v^2_T W \nonumber\\
&\qquad -\frac{2\tilde\beta-1}{2} W^2-\Delta-\mathrm{d}\Delta_T \, ,
\end{align}
where all the terms are given by
\begin{subequations}\label{eq:brol}
	\begin{eqnarray}
	   W&=&\sum_{A \neq T}\frac{\tilde G \tilde m_{A}}{r_{AT}} \, , \label{eq:W}\\
	  W^i&=&\sum_{A \neq T}\frac{\tilde G\tilde m_{A}v_A^i}{r_{AT}} \, , \label{eq:Wi}\\
	  \Delta &=&\sum_{A \neq T} \frac{\tilde G \tilde m_{A}}{r_{AT} }\left[-(\tilde\gamma+1)v_A^2 +\frac{1}{2}\bm r_{AT}.\bm a_A\right. \label{eq:delta}\\
	  &&\qquad\left. +\frac{(\bm r_{AT}.\bm v_A)^2}{2r_{AT}^2}+(2\tilde\beta-1)\sum_{B\neq A}\frac{\tilde G\tilde m_{B}}{r_{AB}}\right]\, , \nonumber
	\end{eqnarray}\\
	\begin{eqnarray}
	    \mathrm{d}\Delta_T&=&2\sum_{A\neq T,B\neq A}\mathrm{d}\tilde \beta^A \frac{\tilde G \tilde m_{A}}{r_{AT}}\frac{\tilde G \tilde m_{B}}{r_{AB}} + \mathrm{d}\tilde \beta^T  W^2 \, , \\
	\tilde\gamma&=&\frac{1 - \alpha_0^2}{1 +\alpha_0^2} =\frac{Z_0-2\left({D^*_{g0}}'- \frac{1}{2}\frac{f'_0}{f_0} \right)^2}{Z_0+2\left({D^*_{g0}}'- \frac{1}{2}\frac{f'_0}{f_0} \right)^2}\, , \label{eq:ppn_gamma}\\
	\tilde\beta-1&=&\frac{\beta_0}{2}\frac{\alpha_0^2}{(1+\alpha_0^2)^2}=\frac{1-\tilde\gamma^2}{8} \beta_0 \, , \label{eq:ppn_beta}\\
	\mathrm{d}\tilde \beta^T&=& \frac{\tilde\beta_{T0}}{2}\frac{\alpha_0^2 }{(1+\alpha_0^2)^2} \approx \frac{\bar\beta_{T0}}{Z_0} \frac{\alpha_0^2 }{(1+\alpha_0^2)^2} \, ,\nonumber\\\label{eq:tilde_beta_A}
	\end{eqnarray}
	with $\tilde G$, $\delta_A$ and $\delta_{AT}$ given by Eqs.~(\ref{eq:deltas}) and $\tilde m_{A}$ given by Eq.~(\ref{eq:tilde_m}). The $\delta_A$ coefficient can now be expressed in term of the $\tilde \gamma$ parameter as
	\begin{align}
	 \delta_A&= \frac{1-\tilde\gamma}{2}\, \frac{\tilde\alpha_{A0}}{\alpha_0} \,. %,\\
	 %
	%\delta_{AT}&= \frac{1-\tilde\gamma}{2} \, \frac{ \tilde\alpha_{A0}\tilde\alpha_{T0}}{\alpha_0^2} \, .
	\end{align}
\end{subequations}
Note that $\tilde\gamma=1$ implies $\delta_A=0$ but not  necessarily $\delta_{AT}=0$. The potentials $W$, $W^i$ and $\Delta$ are standard and the coefficients $\tilde \gamma$ and $\tilde \beta$ can be identified as the standard PPN parameters. One can note that although the parameter $\tilde \gamma_{AT}$ from Eq.~(\ref{eq:gamma_AT}) can be either one, less than one or even more that one \cite{minazzoli:2013fv}, the universal parameter $\tilde \gamma$ can only be less than one or equal to one (in the decoupling limit).

The correction to the Newtonian potential $\tilde W_T$ encode the violation of the universality of free fall. Note however that the $\mathrm{d}\Delta_T$ term appearing at the post-Newtonian level of the Lagrangian is new. This term appears only if there is a non-linear matter-dilaton coupling (see Sec.~\ref{sec:DD_nonlin}) which implies a non-vanishing $\tilde\beta_{A0}$ coefficients as can be seen from equations (\ref{eq:tilde_beta}). This term that is related to the violation of the UFF at the first post-Newtonian order can safely be neglected in most cases. It becomes important only when $\bar\beta_{T0} \alpha_0 /Z_0 \gg \bar \alpha_{T0}$ (which happens for specific coupling functions characterized by $|D_{i0}''| \gg |D_{i0}'|$).

 %-------------------------------------------------------------------------------------

\subsubsection{Simplified equations of motion}
Using the previous simplifications, the equation of motion~(\ref{eq:eqm}) now becomes
\begin{widetext}
\begin{eqnarray}
\bm {a}_T&=&  (1+\delta_T)\bm{\nabla}\tilde W_T +\frac{1}{c^2}\left[\tilde \gamma v_T^2 \bm{\nabla} W-2(\tilde\beta+\tilde\gamma)W \bm \nabla W-\bm \nabla \Delta-\bm \nabla\mathrm{d} \Delta_T -2  (\tilde\gamma+1) \left(\bm{v}_T.\bm{\nabla} W\right)\bm{v}_T\right. \label{eq:eqm_simpl} \\
&&\left.\qquad+2(\tilde\gamma+1)\left(\bm{\nabla}\times\bm{ W}\right)\times\bm{v}_T-(2\tilde\gamma+1)\bm v_T\partial_t W  +2(\tilde\gamma+1)\partial_t \bm { W} \right] \, . \nonumber
\end{eqnarray}
The developed form of these equations are given by
\begin{align}
	\bm {a}_T=&-\sum_{A\neq T} \frac{\tilde G\tilde m_A}{r_{AT}^3}\bm r_{AT}\left(1+\delta_T+\delta_{AT}\right)- \bm Q_T \label{eq:EIH_s} \\
	&-\sum_{A\neq T} \frac{\tilde G\tilde m_A}{r_{AT}^3c^2}\bm r_{AT}\Bigg\{\tilde\gamma v_T^2+(\tilde\gamma+1)v_A^2-2(1+\tilde\gamma)\bm v_A.\bm v_T  -\frac{3}{2}\left(\frac{\bm r_{AT}.\bm v_A}{r_{AT}}\right)-\frac{1}{2}\bm r_{AT}.\bm a_A \nonumber \\
	&\hspace{3cm}-2(\tilde\gamma+\tilde\beta+\mathrm{d}\tilde\beta^T)\sum_{B\neq T}\frac{\tilde G\tilde m_B}{r_{TB}}-(2\tilde\beta+2\mathrm{d}\tilde\beta^A-1)\sum_{B\neq A}\frac{\tilde G\tilde m_B}{r_{AB}}\Bigg\} \nonumber \\
	&+\sum_{A\neq T}\frac{\tilde G\tilde m_A}{c^2r_{AT}}\left[2(1+\tilde\gamma)\bm r_{AT}.\bm v_T-(1+2\tilde\gamma)\bm r_{AT}.\bm v_A\right](\bm v_T-\bm v_A) + (3+4\tilde \gamma)\sum_{A\neq T} \frac{\tilde G \tilde m_A}{c^2r_{AT}}\bm a_A\nonumber \, ,
\end{align}
where $\bm r_{AT}= \bm x_T - \bm x_A$. In this equation, we included a term $\bm Q_T$ that parametrizes a potential correction due to the non-geodesic motion of celestial and artificial bodies. For celestial bodies it could be due, for example, to the interaction of their quadrupole moments with the external masses~\cite{brumberg:1989fk,brumberg:1991uq,brumberg:1992yj,soffel:2003bd}. For spacecrafts it could be due, for instance, to radiation pressure.
\end{widetext}
These are the standard PPN equations of motion (EIH equations of motion) \cite{will:1993fk} with a modification of the Newtonian force to include the violation of the equivalence principle (encoded in $W_{T}$ or in the coefficients $\delta_T$ and $\delta_{AT}$). Post-Newtonian terms also have small differences compared to the standard EIH equations coming from the $\mathrm{d}\Delta_T$ term. These terms are non-negligible only in the case of a non-linear matter-dilaton coupling characterized by $|D_{i0}''| \gg |D_{i0}'|$ (see the discussion in the previous section). 

 %-------------------------------------------------------------------------------------
\subsection{Nordtvedt effect}\label{sec:nordtvedt}
So far, we have neglected hypothetical effects due to the extension and the gravitational self-energy $\Omega_A$ of celestial  bodies since the modeling used is based on a test mass action. The full development in order to study these effects would be to derive the equation of hydrodynamics and reintegrate them over the bodies as it has been done in the standard PPN framework in~\cite{will:1971zr,will:1993fk}. Including a violation of the Einstein equivalence principle in this formalism is beyond the scope of this paper. Nevertheless, a simple heuristic argument allows one to show that the main effect coming from the gravitational self-energy is the usual Nordtvedt effect parametrized by~\cite{nordtvedt:1968uq,*nordtvedt:1968ys,*nordtvedt:1969vn,*nordtvedt:1971mw}
\begin{equation}
 \frac{m^G_A}{m^I_A}=1-\eta_N\frac{|\Omega_A|}{m_Ac^2} \, ,
\end{equation}
where $m^G_A$ refers to the gravitational mass (or passive mass) and $m^I_A$ refers to the inertial mass. This is a consequence of a violation of the Strong Equivalence Principle (SEP). The idea is to introduce the sensitivity of the extended body defined as~\cite{will:1989qy,alsing:2012px}
\begin{equation}
 s_A =-\frac{\partial \ln m_A}{\partial \ln \tilde G} \, ,
\end{equation}
In the weak field limit, the sensitivity of a body is related to its gravitational self-energy $s_A=|\Omega_A|/m_Ac^2$~\cite{eardley:1975ap,zaglauer:1992ap,alsing:2012px}. It is possible to relate the sensitivity of the body to the coupling $\alpha_A(\varphi)$ defined as
\begin{equation}
 \alpha_A=\frac{\partial \ln m_A}{\partial \varphi}=-s_A \frac{\partial \ln \tilde G}{\partial \varphi}=-\frac{|\Omega_A|}{m_Ac^2} \frac{\partial \ln \tilde G}{\partial \varphi} \, .
\end{equation}
This term will be additional to the terms presented in Sec.~\ref{sec:DD} that parametrizes a violation of the Einstein equivalence principle. Therefore, the decomposition (\ref{eq:tilde_alpha}) will have an additional term and becomes
\begin{equation}
 \tilde\alpha_{A0}=\sqrt{\frac{2}{Z_0}}\left[ \bar\alpha_{A0} -\frac{|\Omega_A|}{m_Ac^2} \frac{\partial \ln \tilde G}{\partial \varphi} \right]\, ,
\end{equation}
where the first term is unchanged and where the derivative from the second term can easily be evaluated by using Eq.~(\ref{eq:tildeG}). The parameter $\delta_A$ introduced in  Eq.~(\ref{eq:mg_mi}) as $m^G_A/m^I_A=1+\delta_A$ and whose expression is given by Eq.~(\ref{eq:delta_A}) will now have an aditional term given by
\begin{equation}\label{eq:delta_a_SEP}
 \left[ \delta_A\right]_\textrm{SEP} = - (4\tilde\beta-\tilde\gamma-3)\frac{|\Omega_A|}{m_Ac^2} = -\eta_N \frac{|\Omega_A|}{m_Ac^2}\, ,
\end{equation}
where the $\tilde\gamma$ and $\tilde\beta$ coefficients are the ones defined in Eqs.~(\ref{eq:brol}). This expression shows that the leading contribution of the gravitational self-energy of bodies is parametrized by an additional term in the expression of $\delta_A$ that is parametrized by the Nordtvedt parameter whose value is given by
\begin{equation}
 \eta_N=4\tilde\beta-\tilde\gamma-3\, ,
\end{equation}
which is the standard expression of the Nordtvedt parameter in the PPN formalism~\cite{will:1971zr,will:1993fk}.  As mentioned in the beginning of the section, the heuristic argument presented here allows to derive the main term appearing due to gravitational self-energy but more subtile effects may arise from a more complete hydrodynamical calculation. For the purpose of this work, it is sufficient to consider that gravitational self enery will produce an additional contribution in the expression of $\delta_A$ that takes the form given by Eq.~(\ref{eq:delta_a_SEP}).

 %-------------------------------------------------------------------------------------
\subsection{Temporal evolution of the gravitational constant}\label{sec:Gvaries}
In Sec.~\ref{sec:eom} and \ref{sec:simplifications}, all the quantities with a subscript 0 depend on the background value of the scalar field $\varphi_0$. This background value  usually refers to the cosmological value for the scalar field. This quantity is therefore not strictly constant and will vary with time at cosmological scales. This will produce further effects that may affect the motion of celestial bodies. The question of connecting the Solar System reference frame to the cosmology within the dilaton theory is beyond the scope of this paper. In order to do so, one would have to study the cosmological behavior of the field equations before matching the local reference frame to the cosmological solution. This problem has been considered recently in GR in~\cite{kopeikin:2012dq,*kopeikin:2015fk}. 

Nevertheless, one can phenomenologically work out the main features coming from a temporal variation of the background scalar field. For that purpose, it is sufficient to work with the standard Newtonian equation and keep the leading term produced by a non-vanishing $\dot \phi_0$ (where the dot refers to the temporal derivative in terms of the local time coordinates). At the Newtonian order, starting with the Lagrangian from Eq.~(\ref{eq:Lag_New}) and assuming that $m_{T0}$, $m_{A0}$ and $G_{AT}$ depends on time through a dependance on $\phi_0$, one can easily show that the equations of motion will exhibit an additional term 
\begin{eqnarray}
 \bm a_{T;\dot \varphi_0}&=&-\sum_{A \neq T} \left. \frac{d \ln  (G_{AT} m_{A0})}{dt}\right|_{t_0} \nonumber
 \, (t-t_0)\, \frac{G_{AT}  m_{A0}}{r_{AT}^3}\bm r_{AT}\\
 &&- \frac{d \ln m_{T0}}{dt } \bm v_T  \, ,
\end{eqnarray}
with $t_0$ a reference time and where all time dependent quantities have been evaluated at $t_0$.

Using Eq.~(\ref{eq:gat}) and expressing the gravitational constant in units defined by Eq.~(\ref{eq:gbar}), one gets
\begin{align}
 &\frac{d \ln G_{AT}m_{A0}}{dt}=\left[2\alpha_0 + \tilde \alpha_{A0}+\frac{2\alpha_0\beta_0}{1+\alpha_0^2}\right]\dot \phi_0 \nonumber\\
 &\quad+ \frac{\dot\phi_0}{1+\alpha_0^2}\Bigg[\beta_0(\tilde\alpha_{T0}+\tilde\alpha_{A0})+\alpha_0 (\tilde\beta_{T0}+\tilde\beta_{A0})\Bigg] \, .
\end{align}

Keeping only the leading terms leads to an additional terms of the form
\begin{align}
 \bm a_{T;\dot \varphi_0}\approx& -\dot \phi_0 (t-t_0) \sum_{A \neq T}  (2\alpha_0+\tilde\alpha_{A0}) \frac{\tilde G\tilde m_{A0}}{r_{AT}^3}\bm r_{AT}  \\
 &- \tilde\alpha_{T0}\dot \phi_0\bm v_T \, , \nonumber
\end{align}
that should be added to the standard equations of motion from Eq.~(\ref{eq:EIH_s}).

  %-------------------------------------------------------------------------------------
\section{Propagation of light}\label{sec:light}

In the geometric optic approximation, light follows null geodesics even if there is a multiplicative coupling between the electromagnetic (EM) Lagrangian and the scalar field \cite{minazzoli:2014ao,*minazzoli:2014pb,minazzoli:2013zl,hees:2014pr,*hees:2015gg}. Since the EM Lagrangian is conformally invariant in four dimensions (equivalently since light rays follow null geodesics defined by $ds^2=0$), both the Einstein and the original ``string'' frames can be used to derive the propagation of light, they lead to the same results. For example using Time Transfer Functions formalism~\cite{le-poncin-lafitte:2004cr,*teyssandier:2008nx,*hees:2014nr,*bertone:2014cq,hees:2014fk} at first post-Newtonian order, one can show that the coordinate propagation time is given by
\begin{eqnarray}
 t_R-t_E &=& \frac{R_{ER}}{c}+\frac{R_{ER}}{2c}\int_0^1 \left[h_{00}+N^i_{ER}N^j_{ER}h_{ij}\right]_{z^{\alpha}(\lambda)} d\lambda \nonumber \\
  &&+\mathcal O(1/c^4) \, ,
\end{eqnarray}
where $E$ is the emission event, $R$ is the reception event, $h_{\mu\nu}=g_{\mu\nu}-\eta_{\mu\nu}$ is the post-Minkowksian term of the metric and where the integral is performed on a straight line (parametrized by $z^\alpha(\lambda)$) between the emitter and the receiver. Using any of the metric presented above ($g^*_{\mu\nu}$, $g_{\mu\nu}$ or $\bar g_{\mu\nu}$), one gets
\begin{equation}\label{eq:tr_te}
 t_R-t_E = \frac{R_{ER}}{c}+2\frac{R_{ER}}{c}\int_0^1 w_{0*}\left(z^{\alpha}(\lambda)\right) d\lambda \, .
\end{equation}
Moreover, we have
\begin{equation}
 w_{0*}=\sum_A \frac{G_* m^*_{A0}}{ r_A}=\frac{\tilde \gamma+1}{2}\sum_A \frac{\tilde G\tilde m_{A0}(1-\delta_A)}{r_A} \, .
\end{equation}
Therefore, propagation of light follows geodesic of the standard PPN metric with an additional mass rescaling. For example, the standard Shapiro delay for a single source now writes
\begin{subequations}
 \begin{eqnarray}
 \delta t_\textrm{Shap}&=& \left[1+\tilde\gamma-\delta_A\right] \frac{\tilde G\tilde m_{A0}}{c^2} \nonumber\\
 &&\hspace{1cm}\times\ln \frac{r_E+r_E+r_{ER}}{r_E+r_R-r_{ER}} \label{eq:shapiro}\\
 &=&\left[2+\left(\tilde\gamma-1\right)\left(1+\frac{\tilde\alpha_{A0}}{\alpha_0}\right)\right]\frac{\tilde G\tilde m_{A0}}{c^2} \nonumber \\
  &&\hspace{1cm} \times \ln \frac{r_E+r_E+r_{ER}}{r_E+r_R-r_{ER}} \, ,
\end{eqnarray}
\end{subequations}
which shows that the $\tilde \gamma$ PPN measured with light propagation is in theory slightly different from the one appearing in the equations of motion. Nevertheless, the difference is numerically small since $\tilde\alpha_{A0}/\alpha_0 \ll 1$ and in most practical application, one can simply use the usual PPN Shapiro delay formula.

  %-------------------------------------------------------------------------------------
\section{Observables}\label{sec:obs}
In the two previous sections, we have carefully derived the effects of a massless dilaton on celestial body equations of motion and on the propagation of light. Both of these are know to be coordinate dependent and as such, do not (necessarily) correspond to observable quantities. In order to analyze observations, it is important to rely on observables that are coordinate independent. The fact that we are considering a theory that is breaking the Einstein Equivalence Principle means that the expressions of the observables in terms of the space-time metric are not necessarily similar to their GR expressions. In this section, we will derive different observables (in particular the ones that rely on observations made with accurate clocks) and enlighten the difference between GR and dilatonic theories.

  %-------------------------------------------------------------------------------------
\subsection{Time measured by a specific clock}
\label{sec:PTvsMT}

Because of the violation of the equivalence principle, the time measured by a clock does not correspond necessarily to the proper time $d\tau=-g_{\mu\nu}dx^\mu dx^\nu$ appearing in the action (\ref{eq:smat}). The measured time becomes now dependent on the composition of clocks and on the local value of the scalar field. This is due to the breaking of the equivalence principle and is known as a violation of the Local Position Invariance~\cite{will:1993fk,will:2014la}. This can easily be illustrated by an atomic clock that depends on a specific atomic transition.  For instance, Prestage et al.~\cite{prestage:1995pl} and later on Flambaum and Tedesco \cite{flambaum:2006pr} give the dependency of hyperfine transition based frequencies on fundamental parameters such as the fine structure constant, the mass of the electron or the mass of the proton. Since all of those parameters depend on the value of the dilaton (as mentioned in Sec.~\ref{sec:DD}), frequencies given by those clocks will also depend on the dilaton field. 

In general, let us condider a clock of internal composition $I$ that is delivering a frequency $\nu^I(\varphi)$ that depends on its internal composition and on the value of the scalar field. The time given by this clock can therefore be written as
\begin{equation}
 d\hat \tau_I =  \frac{\nu^I(\varphi_0)}{\nu^I(\varphi)}~d\tau \, ,
\end{equation}
where the first term encodes all the composition dependence of the clocks and the second term is the standard proper time of the space-time metric $g_{\mu\nu}$ that would be observed if no violation of the Einstein Equivalence Principle is considered. Note that this relationship defines a new metric conformally related to the original one $g_{\mu\nu}$ and to the Einstein frame one $g^*_{\mu\nu}$ by
\begin{equation}
 \hat g_{\mu\nu}^I=\left(\frac{\nu^I(\varphi_0)}{\nu^I(\varphi)}\right)^2 g_{\mu\nu} = \left(\frac{\nu^*_I(\varphi_0)}{\nu^*_I(\varphi)}\right)^2 g_{\mu\nu}^* \, ,
\end{equation}
where we introduce the quantity $\nu_I^*=\sqrt{f(\varphi)/f_0}~\nu_I$. Therefore, the clock $I$ will measure the proper time related to this new metric. Similarly to the coefficients $\alpha_A(\varphi)$ introduced in Sec.~\ref{sec:DD}, it is natural to introduce the following quantity
\begin{equation}
 \chi_I(\varphi)= -\frac{\partial \ln \nu^I(\varphi)}{\partial \varphi}\, .
\end{equation}
 This quantity depends on the composition of the clock $I$ and has a similar role to the $\alpha_A(\varphi)$ quantity defined in Eq.~(\ref{eq:alpha}). Therefore,  it is also natural to  introduce the analogous of $\alpha_A^*(\varphi)$ defined by Eq.~(\ref{eq:alpha_s}), such that
\begin{subequations}
 \begin{align}
 \chi_I^*(\phi) &= -\frac{\partial \ln \nu_I^*}{\partial \phi} \\
 &=\sqrt{\frac{2}{Z(\varphi)}}\left(\chi_I(\varphi) - \frac{1}{2} \frac{f'(\varphi)}{f(\varphi)}\right) \, .%= \chi(\phi) + \tilde \chi_I(\phi) \, ,
\end{align}
\end{subequations}
It will be useful to introduce the quantity $\tilde \chi_I(\phi)$ (similar to $\tilde \alpha_A$) defined by
\begin{subequations}
\begin{equation}
	 \chi_I^*(\phi)=\alpha(\phi) + \tilde \chi_I(\phi) \, ,
\end{equation}
and that can be written as
\begin{equation}\label{eq:tilde_chi}
	\tilde \chi_I(\phi) =\sqrt{\frac{2}{Z(\varphi)}}\left[\chi_I(\varphi)-{D_g^*}'(\varphi)\right] \, .
\end{equation}
\end{subequations}

The expression of the measured time is now given by
\begin{equation} \label{eq:driftmeasuredtime}
 d\hat \tau_I=\frac{\nu_I^*(\phi_0)}{\nu_I^*(\phi)}\left(-g_{00}^*-2g_{0i}^*\frac{v^I_I}{c}-g_{ij}^*\frac{v_I^iv_I^j}{c^2}\right)^{1/2} dt \, .
\end{equation}
It turns out that the expression of $d\hat \tau_I/dt$ is almost exactly the same to the expression of the Lagrangian given by Eq.~(\ref{eq:lag_start}) in Sec.~\ref{sec:eom}. Indeed, one only needs to replace $\alpha_T^*$ by $\chi_I^*$. Therefore, the equations for the measured time can be written as follows
\begin{align}
 \frac{d\hat \tau_I}{dt} &=1-\frac{v_I^2}{2c^2} -\frac{\hat W_I}{c^2}\label{eq:dhattau_dt}\\
&+\frac{1}{c^4}\left[-\frac{v^4_I}{8}-v_I^2\left(\frac{\hat W_I}{2} + \hat U_I\right)+4 w_*^j v_I^j +\frac{\hat W_I^2}{2}+\breve \Delta_I\right] \nonumber\, ,%\\
\end{align}
with 
\begin{subequations}
\begin{align}
 \hat W_I&=\sum_A \frac{ G_*  m_{A0} }{r_{AI}}\left(1+\alpha_{A0}^*\chi_{I0}^*\right) \, \\
\hat U_I&=\sum_A \frac{ G_*  m_{A0} }{r_{AI}c^2}\left(1- \alpha_{A0}^*\chi_{I0}^*\right)\, , \\
\breve\Delta_I&= \sum_A \frac{G_* m_{A0}}{r_{AI}}\Bigg\{ -2v_A^2 +\left(1+\alpha_{A0}^*\chi_{I0}^*\right)\nonumber\\
&\times \left[\frac{\bm r_{AI}.\bm a_A}{2}+\frac{1}{2}\left(\frac{\bm r_{AI}.\bm v_A}{r_{AI}}\right)^2+\sum_{B\neq A}\frac{G_{AB}m_{B0}}{r_{AB}}\right] \Bigg\} \nonumber \\
 &+ \frac{1}{2} \left.\frac{\partial \chi^*_{I}}{\partial \phi}\right|_0 \sum_{A,B}\frac{G_* m_{A0}\alpha_{A0}^*}{r_{AI}}\frac{G_* m_{B0}\alpha_{B0}^*}{r_{BI}}  \\
&+ \chi_{I0}^*\sum_A  \beta_{A0}^*\frac{G_* m_{A0}}{r_{AI}}\sum_{B\neq A}\alpha_{B0}^*\frac{G_* m_{B0}}{r_{AB}} \, .\nonumber
\end{align}
\end{subequations}

\subsubsection{Simplification for practical use}
For applications, it is convenient to use a simplified version of Eq.~(\ref{eq:dhattau_dt}). The procedure to simplify this equation is similar to what has been performed in Sec.~\ref{sec:simplifications} and consists to keep the terms violating the equivalence principle only in the $1/c^2$ part. This simplification leads to 
\begin{align}
 \frac{d\hat \tau_I}{dt} &=1-\frac{v_I^2}{2c^2} -\frac{\hat W_I}{c^2}+\frac{1}{c^4}\Bigg[-\frac{v^4_I}{8}-\frac{2\tilde\gamma+1}{2}v_I^ 2W  \nonumber\\
&+2(\tilde \gamma+1) W^j v_I^j +\frac{2\tilde\beta-1}{2} W^2+\Delta +\hat \Delta_I\Bigg]\label{eq:dhattau_dt_simp}\, ,%\\
\end{align}
where $\hat W_I$ is given by 

\begin{equation}
	\hat W_I=\sum_A \frac{\tilde G \tilde m_{A} }{r_{AI}} \left(1+\hat\delta_I+\hat \delta_{IA}\right) \, , \label{eq:hat_Wi} 
\end{equation}
with 
\begin{subequations}\label{eq:hat_deltas}
\begin{align}
	\hat \delta_I&=\frac{\alpha_0 \tilde \chi_{I0}}{1+\alpha_0^2}=\frac{1-\tilde\gamma}{2}\frac{\tilde \chi_{I0}}{\alpha_0} \, ,\\
	\hat \delta_{IA}&=\frac{\tilde \alpha_{A0} \tilde \chi_{I0}}{1+\alpha_0^2} \, ,
\end{align}
\end{subequations}
and the potentials $W$, $W^i$ and $\Delta$ as well as the PPN parameter $\tilde \gamma$ and $\tilde\beta$ are standard and given by Eqs.~(\ref{eq:brol}). Finally, $\hat \Delta_I$ is given by
\begin{subequations}
	\begin{equation}
		\hat \Delta_I=\mathrm{d}\hat\beta_I W^2 +2 \sum_{A,B\neq A} \mathrm{d}\tilde \beta^A \frac{\tilde G \tilde m_{A}}{r_{AI}}\frac{\tilde G\tilde m_B}{r_{AB}} \, ,
	\end{equation}
	with $\mathrm{d}\tilde \beta^A $ given by Eq.~(\ref{eq:tilde_beta_A}) and
	\begin{equation}
		\mathrm{d}\hat\beta_I=\frac{1}{2}\left. \frac{\partial \tilde \chi_I}{\partial \phi}\right|_0\frac{\alpha_0^2}{1+\alpha_0^2}\, ,
	\end{equation}
	where
	\begin{align}
	 \left. \frac{\partial \tilde \chi_I}{\partial \phi}\right|_0=& \frac{2}{Z_0}\left[\chi_{I0}'-{D^*_{g0}}''\right]-\frac{Z'_0}{\sqrt{2}Z_0^{3/2}} \tilde \chi_{I0} \, .
	\end{align}

\end{subequations}

\subsubsection{Expression of the couplings in the case of a clock working on a hyperfine transition}
The coupling $\chi_{I}(\varphi)$ related to a clock $I$ working on an atomic transition is simply given by~\cite{nordtvedt:2002uq,damour:1997qv,*damour:1999fk}
\begin{equation}
	\chi_I(\varphi)=-\frac{\partial \ln E_I(\varphi)}{\partial \varphi}\, ,
\end{equation}
where $E_I$ is the expression of the transition energy as a function of the dilaton field. In the case of a clock working on a hyperfine transition, the expression of the transition energy as a function of fundamental parameters (electron charge, mass of particles, \dots) is given in~\cite{flambaum:2004pr,flambaum:2006pr} and reads
\begin{equation}
%E^I_\textrm{hyperfine} \propto (m_e \af^2) g_I \frac{m_e}{m_p} \af^2 F_\textrm{rel}(Z\af)\, , \label{eq:reffreqclockF}
E^I_\textrm{hyperfine} \propto \frac{m_e^2}{m_p} \af^{4+K_{\textrm{rel}}} \left(\frac{\hat{m}}{\Lambda_3} \right)^{\kappa_q} \left(\frac{m_s}{\Lambda_3} \right)^{\kappa_s}, \label{eq:reffreqclockF}
\end{equation}
where $m_e$ and $m_p$ are the electron and proton mass respectively, $\hat m$ is the mean value of the light quark masses (up and down), $m_s$ is the mass of the strange quark and $\af$ is the fine structure constant. $K_{\textrm{rel}}$ comes from the Casimir factor and reads (at the $s$-wave approximation\footnote{Numerical many-body calculations give more accurate results \cite{dzuba:1999pz}.}) $K_{\textrm{rel}}= (Z \af)^2 (12 \lambda^2-1)/(\lambda^2 (4 \lambda^2-1))^{-1}$, where $\lambda=[1-(\af  Z)^2]^{1/2}$, while $\kappa_q$ and $\kappa_s$ come from the nuclear magnetic moment and are computed in \cite{flambaum:2006pr}. Until now, since we used Damour and Donoghue microphysical model \cite{damour:2010zr,*damour:2010ve}, we neglected possible contributions from strange quarks. In order to be coherent with the rest of the paper we shall continue to neglect such contribution in the following. Their contribution is nevertheless only one order of magnitude smaller than the contribution from light quarks \cite{flambaum:2006pr}. The derivative of the previous formula gives
\begin{subequations}\label{eq:chi_param}
\begin{align}
%	\chi_I(\varphi)&=h'(\varphi) \Big[-2 d_{m_e}-4 d_e -d_e L_d F_\textrm{rel}(Z\af) \nonumber\\
%	 &\hspace{2cm}+ d_g^* + \bar \alpha_p \Big]\, , 
%\chi_I(\varphi)&= {D_g^*}'(\varphi) + \bar \alpha_p(\varphi)-2 D'_{m_e}(\varphi)-(4+K_{\textrm{rel}})D_e'(\varphi), \nonumber\\
%	 &\hspace{2cm} - \kappa_q (D_{\hat m}'(\varphi)-D'_g(\varphi))\, , 
\chi_I(\varphi)&= \alpha_p(\varphi)-2 D'_{m_e}(\varphi)-(4+K_{\textrm{rel}})D_e'(\varphi), \nonumber\\
	 &\hspace{2cm} - \kappa_q (D_{\hat m}'(\varphi)-D'_g(\varphi))\, , 
\end{align}
where the functions $D_i(\varphi)$ are the dilaton-matter coupling functions introduced in Sec.~\ref{sec:DD}.\footnote{Let us remind that in a linear coupling model like the one from \cite{damour:2010zr}, one has $D_i'(\varphi)=d_i$.)}  Therefore, one may have to consider strange quarks' effects in future analysis. $\alpha_p$ is related to the proton composition and is given by \footnote{The difference with the numerical application of Eqs. (\ref{eq:alpha_NL}) lies in the fact that Eqs. (\ref{eq:alpha_NL}) is an approximation that is not suitable for protons. See Appendix \ref{app:alphacoeff}.} (see Appendix \ref{app:alphacoeff})
\begin{eqnarray}
	\alpha_p(\varphi)&=&D'_g(\varphi)+0.037  (D'_{\hat m}(\varphi)-D'_g(\varphi)) \label{eq:alpha_p}\\
	 &&- 0.0017 (D'_{\delta m}(\varphi)-D'_g(\varphi)) \nonumber\\
	&&+5.5\times10^{-4} (D'_{m_e}(\varphi)-D'_g(\varphi))\nonumber\\
	&&+6.8\times10^{-4} D'_e(\varphi) \nonumber \,.
\end{eqnarray}

 Introducing this expression in Eq.~(\ref{eq:tilde_chi}) gives
\begin{align}
	\tilde \chi_I(\phi) &=  \sqrt{\frac{2}{Z(\varphi)}}\Bigg[ \bar\alpha_p -2 D'_{m_e}(\varphi) \label{eq:tilde_chi_param} \\
	&-(4+K_{\textrm{rel}}) D_e'(\varphi)-  \kappa_q (D_{\hat m}'(\varphi)-D'_g (\varphi))\Bigg] \, , \nonumber 
\end{align}
where $\bar \alpha_p$ is given by
\begin{eqnarray} \label{eq:alphabarp}
	\bar \alpha_p(\varphi)&=&\alpha_P-{D_g^*}'(\varphi)\\
	&=&-0.056  (D'_{\hat m}(\varphi)-D'_g(\varphi)) \nonumber\\
	 &&- 0.0017 (D'_{\delta m}(\varphi)-D'_g(\varphi)) \nonumber\\
	&&+5.5\times10^{-4} (D'_{m_e}(\varphi)-D'_g(\varphi))\nonumber\\
	&&+4.1\times10^{-4} D'_e(\varphi) \,. \nonumber
\end{eqnarray}

\end{subequations}

   %-------------------------------------------------------------------------------------

 %-------------------------------------------------------------------------------------
 \subsection{Range}\label{sec:range}
 
The range is obtained as a difference of measured time given by two clocks (or even one in special configurations like when light is reflected before going back to where it originated, such as in lunar laser ranging for instance \cite{turyshev:2013ex}). let us assume that the signal is emitted by a clock $\mathcal C_E$ of composition $E$ characterized by $\chi_E$. Similarly, the signal is received by a clock $\mathcal C_R$ whose composition is characterized by $\chi_R$. The measured time provided by $\mathcal C_E$ and $\mathcal C_R$ are given by $\hat \tau_E$ and $\hat \tau_R$ which are given by the integration of Eq.~(\ref{eq:dhattau_dt_simp}). In general, the range is therefore given by
\begin{equation}
	\mathcal R=\hat \tau_R(t_R)-\hat \tau_E(t_E) \, ,
\end{equation}
where $t_R$ and $t_E$ are the coordinate reception and emission time of the electromagnetic signal and are related through Eq.~(\ref{eq:tr_te}).
 
let us now consider a particular situation where the same type of clock $\mathcal C$ is used at the emission and at the reception and where the EM signal is doing a round trip\footnote{It could be for example the tracking of spacecrafts measured with a ``DSN station-spacecraft-DSN station'' link.}.  In this case, the range $\mathcal R$ is given by
\begin{eqnarray}
 \mathcal R&=&\hat \tau_C (t_R) -\hat \tau_C (t_E) \nonumber \\
 &=&\hat \tau_C (t_R) -\hat \tau_C (t_R) - (t_E-t_R) \frac{ d\hat \tau_C}{dt} \nonumber \\
 &=& (t_R-t_E) - \frac{2R_{ER}}{c^3}\left(\hat W_C + \frac{v_C^2}{2}\right) \nonumber \\
 &=& \mathcal R_\textrm{PPN} - \frac{2R_{ER}}{c^3}\sum_A \frac{\tilde G \tilde m_{A}}{r_{AC}}\left(\hat\delta_C + \hat\delta_{CA}\right) \, , \label{eq:range}
\end{eqnarray}
where $\mathcal R_\textrm{PPN}$ is the expression of the range in the standard PPN formalism and where the $\hat\delta_C$ and $\hat\delta_{CA}$ are depending on the constitution of the clock $\mathcal C$ through Eq.~(\ref{eq:hat_deltas}).

  %-------------------------------------------------------------------------------------
 \subsection{Doppler}
 let us now focus on the frequency shift. Following the methodolgy of~\cite{blanchet:2001ud}, we can write the frequency shift as
\begin{equation}
 \frac{\nu_R}{\nu_E}=\frac{d\hat \tau_E}{d\hat\tau_R}=\left.\frac{d\hat \tau_E}{dt}\right|_{t=t_E} \left(\left.\frac{d\hat \tau_R}{dt}\right|_{t=t_R}\right)^{-1} \frac{dt_E}{dt_R} \, ,
\end{equation}
where the expression of the $d\hat \tau/dt$ are given by Eq.~(\ref{eq:dhattau_dt_simp}) and the $dt_E/dt_R$ can be computed from the PPN metric as usual (see Sec.~\ref{sec:light}). This gives
\begin{eqnarray}
 \frac{\nu_R}{\nu_E}&=&\frac{1 -\frac{\hat W_E}{c^2} - \frac{v^2_E}{2 c^2}}{1 -\frac{\hat W_R}{c^2} - \frac{v^2_R}{2 c^2}} \times \frac{q_R}{q_E}\label{eq:doppler}\\
  &=&\frac{1 -\frac{W_E}{c^2} - \frac{v^2_E}{2 c^2}-\sum_A  \frac{\tilde G \tilde m_A}{c^2 r_{AE}}\left(\hat\delta_E + \hat\delta_{EA}\right) }{1 -\frac{W_R}{c^2} - \frac{v^2_R}{2 c^2}-\sum_A \frac{\tilde G \tilde m_{A}}{c^2r_{AR}} \left(\hat\delta_R+\hat\delta_{RA}\right)} \times \frac{q_R}{q_E} \, , \nonumber
\end{eqnarray}
where $W_{E/R}$ are the standard Newtonian potential at the emission/reception events and the ratio $q_R/q_E$ is due to corrections arising from the transmission of the signal between the two clocks. It includes standard Doppler corrections as well as corrections related to the Shapiro delay. This term is computed from the PPN metric and is given for instance in~\cite{blanchet:2001ud,hees:2014fk}.

  %-------------------------------------------------------------------------------------
\subsection{Angular separation}\label{sec:astrometric}
In general, astrometric observations measure the angular separation between two celestial bodies or stars. The observable in this case is defined as the scalar product of the two wave vectors at the event of reception (see for example~\cite{brumberg:1991uq}). Let us denote by $k^\mu$ and $(k')^\mu$ be the two wave vectors tangent to the two light rays. The angular separation $\theta$ between these two light rays is given by~\cite{teyssandier:2006fk,hees:2014fk,hees:2015rm}
\begin{widetext}
\begin{equation}
	\sin^2\frac{\theta}{2}=-\frac{1}{4}\left[\frac{\left(g^*_{00}+2g_{0k}^*v^k+g_{kl}^*v^kv^l\right)g^{ij}_*(\hat k_i-\hat k_i')(\hat k_j-\hat k_j')}{(1+\hat k_m v^m/c)(1+\hat k_l' v^l/c)}\right]_R \, ,
\end{equation}
\end{widetext}
where the bracket is evaluated at the reception event and where $\hat k_i=k_i/k_0$. This expression is clearly conformally invariant. Following the discussion from Sec.~\ref{sec:light}, the $\hat k_i$ quantities can be computed from the PPN metric (or to be rigourous from the PPN metric with rescaled masses $\tilde m_A  (1-\delta_A)$). After introducing the expression of the metric, one finds
\begin{equation}
		\sin^2\frac{\theta}{2}=\frac{1}{4}\left[\frac{ \left(1-2(1+\tilde\gamma) \frac{W_R}{c^2}-\frac{v_R^2}{c^2}\right) | \hat {\bm k}-\hat {\bm k}' |^2}{(1+\hat{ \bm k}.\bm v_R/c)(1+\hat{\bm k}'.\bm  v_R/c)}\right] \, .
\end{equation}
This expression is exactly the one from the PPN metric~\cite{hees:2014fk}, which shows that the reduction of astrometric observables can safely be done using the standard PPN formalism.

  %-------------------------------------------------------------------------------------
\section{Applications}\label{sec:use}
The goal of this section is to apply all the relations developed above to different applications and is devoted to data analysis. All the post-Newtonian phenomenology developed so far depends on a a large numbers of parameters: $f_0$, $f'_0$, $f''_0$, $\omega_0$, $\omega_0'$, $\omega_0''$, $\dot \phi_0$ and the 5 matter-dilaton coupling functions $D_{i0}$ and their derivatives. It is illusory to think that weak field observations are able to determine all of these parameters. 

For data analysis, it is much more appropriate to consider the following combinations of parameters:
\begin{itemize}
	\item  first, two universal parameters
\begin{subequations}\label{eq:param_fond}
	\begin{align}
		\alpha_0&=\sqrt{\frac{2}{Z_0}}\left[{D_{g0}^*}'-\frac{1}{2}\frac{f'_0}{f_0}\right] \, , \\
		\beta_0&=\frac{2}{Z_0}\left[{D_{g0}^*}''-\frac{1}{2}\frac{f''_0}{f_0}+\frac{1}{2}\left(\frac{f'_0}{f_0}\right)^2\right]-\frac{Z'_0}{Z_0^{3/2}}\frac{\alpha_0}{\sqrt{2}} \, .
	\end{align}
\item	In addition to these two parameters, one needs to introduce the following parameters for each bodies (or for each type of bodies only in order to simplify the problem)
	\begin{align}\label{eq:S_tilde_alpha}
		\tilde \alpha_{A0}&=\sqrt{\frac{2}{Z_0}} \bar \alpha_{A0} , \\
		\tilde \beta_{A0}&= \frac{2}{Z_0}\bar\beta_{A0} -\frac{Z'_0}{\sqrt{2}Z_0^{3/2}} \tilde \alpha_{A0} \, ,
	\end{align}
recalling that $\bar \alpha_{A}$ is given by Eq.~(\ref{eq:bar_alpha_A}) and $\bar\beta_A$ by Eq.~(\ref{eq:bar_beta}). \item Finally, one needs to consider one additional parameter $\tilde \chi_{I0}$ by type of clocks used to make observations. In the case of an atomic clock working on a hyperfine transition, this parameter is related to the fundamental parameter through
\begin{align}		\tilde \chi_{I0} &=  \sqrt{\frac{2}{Z_0}}\Bigg[\bar \alpha_{p0} -2 D'_{m_e 0} \label{eq:S_tilde_chi} \\
	&-(4+K_{\textrm{rel}}) D_{e0}'-  \kappa_q ((D_{\hat m 0}'-D'_{g 0})\Bigg] \, , \nonumber 
\end{align}
where $\bar \alpha_p$ is given in Eq. (\ref{eq:alphabarp}), while $K_\textrm{rel}$ and $\kappa_q$ are given in~\cite{flambaum:2006pr}.
\end{subequations}
\end{itemize}

From these parameters, it is possible to compute the universal PPN parameters
\begin{subequations}\label{eq:param_sum}
	\begin{align}
		\tilde\gamma&=\frac{1-\alpha_0^2}{1+\alpha_0^2}\, ,\\
		\tilde\beta&=1+\frac{\alpha_0^2}{(1+\alpha_0^2)^2}\beta_0 \, ,
	\end{align}
	but also the following parameters that are specific for each body and that parametrize a violation of the equivalence principle
	\begin{align}
	\delta_A&= \frac{\alpha_0\tilde\alpha_{A0}}{1+\alpha_0^2}-(4\tilde\beta-\tilde\gamma-3)\frac{|\Omega_A|}{m_Ac^2}\, , \\
		\delta_{AT}&=\frac{\tilde\alpha_{T0}\tilde\alpha_{A0}}{1+\alpha_0^2}\, , \\
		\mathrm{d}\tilde\beta^A&=\frac{\alpha_0^2}{(1+\alpha_0^2)^2}\tilde \beta_{A0} \, ,
	\end{align}
	where the second part of $\delta_A$ parametrize a violation of the Strong Equivalence Principle (Nordtvedt effect, see Sec.~\ref{sec:nordtvedt}). Finally, the following parameters related to clocks can also be computed
	\begin{align}
		\hat\delta_I&=\frac{\alpha_0\tilde\chi_{I0}}{1+\alpha_0^2}\, ,\\
		\hat\delta_{IA}&=\frac{\tilde\alpha_{A0}\tilde\chi_{I0}}{1+\alpha_0^2}\, ,\\
		\mathrm{d}\hat \beta_I&= \frac{\alpha_0^2}{(1+\alpha_0^2)^2}\tilde\chi_{I0}' \, .
	\end{align}
\end{subequations}
All the quantities defined in Eqs.~(\ref{eq:param_sum}) are sufficient to analyze weak field data. The idea is to use parameters from Eqs.~(\ref{eq:param_fond}) as fundamental parameters in the data analysis and to constrain them from diverse observations involving quantities that are developed in Eqs.~(\ref{eq:param_sum}).
%----------------------------------------------
\subsection{Universality of free fall in the lab}\label{sec:UFF}
One way to test the universality of free fall is to measure the differential acceleration between two test bodies in the gravitational field of another one (typically the Earth). These tests have currently reached a level of accuracy of the order $10^{-13}$~\cite{schlamminger:2008zr}. Let us denote by ``E'' the source of the gravitational field and by ``B'' and ``C'' the two test masses. From Eq.~(\ref{eq:eqm_simpl}),  a test particle B evolving in the gravitational field generated by the body ``E'' will experience an acceleration $\bm {a}_B= \bm {\nabla} W (1+ \delta_{B}+\delta_{EB})$ at the Newtonian level. Therefore the relative acceleration of two particles falling in the field of the body E reads
\begin{eqnarray}
\eta&=&\left(\frac{\Delta a}{a} \right)_{BC} \equiv 2 \frac{|\bm{a}_B-\bm{a}_C|}{|\bm{a}_B+\bm{a}_C|}\approx  |\delta_B-\delta_C+\delta_{EB}-\delta_{EC}| \nonumber\\
&=&  \left| \frac{\alpha_0 (\tilde{\alpha}_{B0}-\tilde{\alpha}_{C0})}{1+\alpha_0^2}+\frac{\tilde\alpha_{E0} (\tilde{\alpha}_{B0}-\tilde{\alpha}_{C0})}{1+\alpha_0^2}\right| \label{eq:eta_UFF}\\
&=&\frac{1-\tilde\gamma}{2} \left|\frac{\bar{\alpha}_{B0}-\bar{\alpha}_{C0}}{\alpha_0}\right| +\frac{1+\tilde\gamma}{2} \left|\tilde\alpha_{E0} (\tilde{\alpha}_{B0}-\tilde{\alpha}_{C0})\right|\, .\nonumber
\end{eqnarray}
This first term in this equation is similar to the one found in \cite{damour:2010zr,*damour:2010ve}. In general, this first term is the dominant one. Nevertheless, there exists a decoupling limit (see Sec.~\ref{sec:decoupling}) when $\alpha_0$ vanishes. In this case, the first term vanishes and only the second term plays a role in the violation of the universality of free fall.

As usual in scalar-tensor theories, a decoupling occurs when $Z_0 \rightarrow \infty$, which corresponds to $\alpha_0\rightarrow 0$ and $\tilde \alpha_{A0} \rightarrow 0$ as can be seen from Eq.~(\ref{eq:alpha_s}). In this limit, equivalence principle violations tend to become unobservable. Also, let us remind that for a family of coupling functions, $Z$ is dynamically driven toward infinity during the evolution of the universe \cite{damour:1994fk,*damour:1994uq,damour:1993uq,*damour:1993kx,gerard:1995fk,*gerard:1997nr,serna:2002ys,jarv:2012vn,minazzoli:2014ao,minazzoli:2014pb}.

%----------------------------------------------
\subsection{Comparison of two different atomic clocks at the same position}\label{sec:compclocks}
An interesting experiment consists in asserting the ratio of the frequencies delivered by two different types of clocks located at the same position $B$. In this kind of experiment, the ratio of the frequencies of the two clocks depends only on the value of the scalar field. The observable is therefore given by
\begin{subequations}\label{eq:freq_pot_all}
\begin{align}
\frac{\nu_I(\varphi_B)}{\nu_J(\varphi_B)}&= \frac{\nu_I(\varphi_0)}{\nu_J(\varphi_0)}\left[1+\left.\frac{\partial \ln \nu_I/\nu_J}{\partial \varphi}\right|_0 (\varphi_B-\varphi_0)\right] \nonumber\\
&= \frac{\nu_I(\varphi_0)}{\nu_J(\varphi_0)}\left[1+\frac{\hat W_I-\hat W_J}{c^2}\right] \label{eq:freq_pot}\\
&=\frac{\nu_I(\varphi_0)}{\nu_J(\varphi_0)}\left[1+\sum_A \frac{\tilde G\tilde m_A}{c^2 r_{AB}}\left(\hat \delta_I-\hat\delta_J +\hat\delta_{IA}-\hat\delta_{JA}\right) \right]\, \label{eq:freq_pot2} .
\end{align}
\end{subequations}
The term $\frac{\nu_I(\varphi_0)}{\nu_J(\varphi_0)}$ is constant and unobservable as long as $\varphi_0$ is considered  constant. It plays a role when one considers a temporal variation due to cosmological evolution of the scalar field (see Sec.~\ref{sec:freq_cosmo}). Different temporal signatures can be used to constrain the dilatonic couplings: a linear drift due to a cosmological evolution of the scalar field which is encoded in the $\nu_{I}(\varphi_0)/\nu_J(\varphi_0)$ terms or some periodical variations due to the evolution of the potentials $\hat W_J$ and $\hat W_I$ (typically annual and sidereal frequencies).

\subsubsection{Other parametrization used in data analysis}
Instead of using the frequency parametrization given by Eqs.~(\ref{eq:chi_param}), it is common to parametrize the ratio between two atomic frequencies by using three constants~\cite{guena:2012ys}: $k_\alpha$, $k_\mu$ and $k_q$ related to the fine structure constant $\af$, to $\mu=m_e/m_p$ and to the ratio $\hat m/\Lambda_3$. A fractional variation of any atomic frequency therefore writes
\begin{equation}\label{eq:another_param}
 d \ln \frac{\nu_I}{\nu_J} = k_\alpha d \ln \alpha_{EM} + k_\mu d \ln \mu + k_q d \ln (\hat m/\Lambda_3) \, ,
\end{equation}
where the three coefficients $k_i$ depend on the atomic frequencies considered and are given by atomic and nuclear structure calculations~\cite{flambaum:2006pr,dinh:2009fk,dzuba:2008uq,flambaum:2009kx}.

Using this parametrization, one finds 
\begin{align}
 \frac{\partial \ln \nu_I/\nu_J }{\partial \varphi}=&\chi_J(\varphi)-\chi_I(\varphi)=k_\mu \left(D_{m_e}'(\varphi)-\alpha_p(\varphi)\right)\nonumber \\
 &+  k_\alpha D'_e(\varphi)+k_g\left(D'_{\hat m}(\varphi)-D_g'(\varphi)\right) \, ,\label{eq:param_nu_2}
\end{align}
with $\alpha_p$ given by Eq.~(\ref{eq:alpha_p}).
This quantity appears directly in Eq.~(\ref{eq:freq_pot}) since
\begin{subequations}\label{eq:diff_delta}
 \begin{align}
  \hat\delta_I-\hat\delta_J=& \sqrt{\frac{2}{Z_0}}\frac{\alpha_0}{1+\alpha_0^2} \left[\chi_{I0}-\chi_{J0}\right] \, , \\
  \hat\delta_{IA}-\hat\delta_{JA}=& \sqrt{\frac{2}{Z_0}}\frac{\tilde\alpha_{A0}}{1+\alpha_0^2} \left[\chi_{I0}-\chi_{J0}\right] \, . 
 \end{align}
\end{subequations}

\subsubsection{Periodical evolution of the frequencies ratio}\label{sec:comp_clock_pot}
Any variation of the ratio of two atomic frequencies can be related to the variation of the gravitational potential as shown by Eq.~(\ref{eq:freq_pot}). Therefore, any periodic modulation of the gravitational potential due to orbital motion or to Earth rotation can be searched in measurements and used to constrain the dilaton/matter couplings by using Eq.~(\ref{eq:freq_pot2}).

For instance, assuming that the gravitational potential of the Sun $W_\odot$ is the dominant gravitational source, one gets from Eq.~(\ref{eq:freq_pot2})
\begin{equation}
\delta \left(\ln \frac{\nu_I}{\nu_J}\right) = \left(\hat \delta_I-\hat\delta_J +\hat\delta_{I\odot}-\hat\delta_{J\odot}\right) \frac{\delta W_\odot}{c^2}\, ,
\end{equation}
where one can either use the definition of the $\hat \delta$ coefficients from Eqs.~(\ref{eq:hat_deltas}) and~(\ref{eq:chi_param}) or use the expressions~(\ref{eq:diff_delta}) with Eq.~(\ref{eq:param_nu_2}).

As the Earth moves on an ellipse around the sun, the gravitational field of the Sun varies on Earth such that one should expect annual variations of the frequency ratios. Several null experiments have put constraints on such variations \cite{guena:2012ys,tobar:2013pr,leefer:2013pl}. Using the parametrization introduced in Eq.~(\ref{eq:another_param}) and a combination of observations of different types of frequencies ratios, it has been possible to produce individual constraints on the  variation of the fine structure constant $\af$, of $\mu=m_e/m_p$ and of $\hat m/\Lambda_3$. These constraints are very easily linked to the dilaton/matter couplings
\begin{subequations}
	\begin{align}
		\frac{d\ln\af}{dW_\odot}&=-\sqrt{\frac{2}{Z_0}}\frac{\alpha_0+\tilde\alpha_\odot}{1+\alpha_0^2} D'_{e0} \nonumber\\
		&=(0.32\pm 0.46)\times 10^{-6}  \, , \\
		\frac{d\ln \mu}{dW_\odot}&=\sqrt{\frac{2}{Z_0}}\frac{\alpha_0+\tilde\alpha_\odot}{1+\alpha_0^2}(\alpha_{p0}-D'_{m_e0})\nonumber\\
		&=(-0.23\pm 2.0)\times 10^{-6}  \, , \\
		\frac{d\ln(\hat m/\Lambda_3)}{dW_\odot}&=\sqrt{\frac{2}{Z_0}}\frac{\alpha_0+\tilde\alpha_\odot}{1+\alpha_0^2}(D'_{g0}-D'_{\hat m 0})\nonumber\\
		&=(-3.07\pm 5.58)\times 10^{-6}  \, ,
	\end{align}
	where the variations given here are with respect to the gravitational potential $W_\odot$. The numerical values are taken from \cite{bize:2015mo}.
\end{subequations}

In particular, it is interesting to mention that these combinations are different from the ones appearing in UFF experiments. In particular, the coupling $D'_{e0}$ that influences very weakly UFF experiments has a direct influence in the comparison of clocks. Moreover, while the coupling $D'_{m_e0}$ does not strongly influence UFF experiments, it is always constrained by comparison of clocks.

%Similarly to UFF experiments, in most cases,  $\hat \delta_J \gg \hat \delta_{JE}$. Nevertheless, close to the decoupling limit (that leads to  $\hat \delta_J=0$ ), the  $\hat \delta_{JE}$ coefficients become crucial.

%----------------------------------------------
\subsubsection{Cosmological and astrophysical effects}\label{sec:freq_cosmo}
As already mentioned in Sec. \ref{sec:Gvaries}, the background value of the dilaton field $\varphi_0$ can actually vary on astrophysical and cosmological scales. In particular, it can vary in time depending on the cosmological evolution of the dilaton field (the cosmological evolution of the dilaton field has been explored for example in \cite{damour:1994fk,damour:2002ys,minazzoli:2014ao,*minazzoli:2014pb}).  Therefore, it is also interesting to search for other variations in the ratio of the two clocks' frequency $\delta \ln (\nu_I/\nu_J)$. Usually, the community is  searching and constraining linear drifts in the frequency ratio~\cite{guena:2012ys,tobar:2013pr,leefer:2013pl,marion:2003zr,*bize:2003ly,*godun:2014pl,*huntemann:2014pl,*fischer:2004ve,*bize:2005pt}. This drift comes from the $\nu_I(\varphi_0)/\nu_J(\varphi_0)$ term of Eqs.~(\ref{eq:freq_pot_all}) and writes
\begin{equation}
	\frac{d \ln \nu_I/\nu_J}{dt}=(\chi_{J0}-\chi_{I0})\dot\varphi_0=(\tilde\chi_{J0}-\tilde\chi_{I0})\dot\phi_0 \, ,
\end{equation}
where the difference of the coefficients $\chi_{I0}$ is given by Eq.~(\ref{eq:param_nu_2}) or the $\tilde\chi_{I0}$ coefficients are given by Eqs.~(\ref{eq:chi_param}). 

Using the parametrization introduced in Eq.~(\ref{eq:another_param}) and a combination of observations of different types of frequencies ratios, it has been possible to produce individual constraints on the temporal variation of the fine structure constant $\af$, of $\mu=m_p/m_e$ and of $\hat m/\Lambda_3$. These constraints are very easily linked to the dilaton/matter couplings
\begin{subequations}
	\begin{align}
		\frac{d\ln\af}{dt}&=D'_{e0}\dot\varphi_0 \nonumber\\
		&=(-0.24\pm 0.23)\times 10^{-16} \textrm{yr}^{-1} \, , \\
		\frac{d\ln \mu}{dt}&=(\alpha_{p0}-D'_{m_e0})\dot\varphi_0\nonumber\\
		&=(1.11\pm 1.39)\times 10^{-16} \textrm{yr}^{-1} \, , \\
		\frac{d\ln(\hat m/\Lambda_3)}{dt}&=(D'_{\hat m 0}-D'_{g0})\dot\varphi_0\nonumber\\
		&=(58.5\pm 29.5)\times 10^{-16} \textrm{yr}^{-1} \, ,
	\end{align}
\end{subequations}
where the numerical values are taken from \cite{bize:2015mo}. However, in order to make a connection between those constraints and the cosmological evolution, one would still need to match the cosmological time with the local coordinate time. This difficult task is out of the scope of the present paper (see also the discussion in Sec.~\ref{sec:Gvaries}).

In addition, it has recently been proposed  to search for oscillations with arbitrary frequencies in models where the dilaton plays also the role of Dark Matter~\cite{arvanitaki:2015pd,vantilburg:2015pl} (in these papers, a potential is nevertheless introduced).

%----------------------------------------------
\subsection{Test of the gravitational redshift}
A measure of the gravitational redshift between two clocks located in two different gravitational field is also considered as a test of the Local Position Invariance, a facet of the Einstein Equivalence Principle~\cite{will:1993fk,will:2014la}. The best current test has been performed by comparing a clock in a rocket with a clock on ground~\cite{vessot:1979fk,*vessot:1980fk} and constraint the gravitational redshift at the level of $10^{-4}$. Several propositions have been made to improve this test using GNSS Galileo 5 and 6 satellites~\cite{delva:2015fk} or with the Atomic Clock Ensemble in Space (ACES) experiment~\cite{cacciapuoti:2011ve}.

Using Eq.~(\ref{eq:doppler}), we can write the gravitational redshift as
\begin{equation}\label{eq:grav_redshift}
 \left.\frac{\Delta \nu}{\nu}\right|_\textrm{grav}=\left.\frac{\nu_R-\nu_E}{\nu_E}\right|_\textrm{grav}=\frac{\hat W_R-\hat W_E }{c^2}\, ,
\end{equation}
where $\hat W$ is the modified Newtonian potential introduced in Eq.~(\ref{eq:hat_Wi}). Usually, deviation from GR is parametrized by a dimensionless parameter $\Upsilon$,\footnote{In Will's formalism~\cite{will:1993fk,will:2014la}, it is written as $\alpha$ but since we already have a lot of $\alpha$ in this document, we decided to use another notation.} such that 
\begin{equation}
 \left.\frac{\Delta \nu}{\nu}\right|_\textrm{grav}=(1+\Upsilon)\frac{\Delta W}{c^2} \, ,
\end{equation}
with $\Delta W= W_R- W_E$ the difference of the Newtonian potentials as experienced by the emission and reception clocks. But it has to be stressed that this parametrization is not ideal in the case of the dilaton theories since it does not parametrize Eq.~(\ref{eq:grav_redshift}) in the general case. 

Nevertheless, it is possible to make the link between the two formalisms for some specific situations. Let us consider a clock orbiting Earth whose frequency  will be compared to the frequency of a clock located on Earth (this corresponds to the tests proposed with the Galilelo 5 and 6 spacecrafts~\cite{delva:2015fk} and to ACES~\cite{cacciapuoti:2011ve}). In that particular case, the gravitational redshift becomes
\begin{align}\label{eq:redgrav}
 \left.\frac{\Delta \nu}{\nu}\right|_\textrm{grav}=\frac{\tilde G\tilde m_\oplus}{r_\oplus} \left(1+\hat\delta_R +\hat \delta_{R\oplus}\right)-\frac{\tilde G\tilde m_\oplus}{r_E} \left(1+\hat\delta_E +\hat \delta_{E\oplus}\right) \, ,
\end{align}
where $\tilde m_\oplus$ is the Earth mass, the subscript $R$ refers to the receiving clock which is located on Earth and $E$ refers to the emitting clock which is orbiting on an elliptical orbit. For the test proposed with Galilelo spacecrafts \cite{delva:2015fk}, only the variable part of the gravitational redshift is used to constrain deviations from GR, the constant part is not observed. In this case, the dependance on the receiving clock disappears and one can safely identify the estimated $\Upsilon$ with $\hat\delta_E+\hat\delta_{E\oplus}$. On the other hand, when the clock PHARAO on ACES will be compared to similar clocks on the ground, then $1+\hat\delta_E +\hat \delta_{E\oplus}$ can be factorized out in Eq. (\ref{eq:redgrav}) and one also recovers $\Upsilon=\hat\delta_E+\hat\delta_{E\oplus}$.

Using the Galileo satellites, we expect to estimate the $\Upsilon$ parameter with an accuracy at the level of $3\times 10^{-5}$~\cite{delva:2015fk} while ACES should reach an accuracy at the level of $10^{-6}$. The estimations can directly be translated into constraints on the matter/dilaton couplings using Eqs.~(\ref{eq:hat_deltas}) and (\ref{eq:S_tilde_chi}). Therefore, the order of magnitude of the constraints on the dilaton/matter couplings that will be reached with ACES are similar to the ones obtained by comparison of two clocks in the lab (see Sec.~\ref{sec:comp_clock_pot}).

On the overall, constraints from clocks comparison are weaker than the ones from universality of free fall. However, one has to stress that they constrain another combination of the fundamental parameters. Such combination is given by Eq.~(\ref{eq:S_tilde_chi}) while UFF experiments give access to the combination in Eq.~(\ref{eq:S_tilde_alpha}). One can see that experiments involving clocks are highly sensitive to $D'_{m_e0}$ and $D'_{e0}$; while UFF experiments are more sensitive to $D'_{g0}$ and to $D'_{m_u0}$ and $D'_{m_d0}$ (see also the discussion in the conclusion of~\cite{wolf:2015fk}). Combining different type of experiments gives a way to disentangle the fundamental parameters $D'_{i0}$ that appear in the combinations (\ref{eq:S_tilde_alpha}) and (\ref{eq:S_tilde_chi}).

%----------------------------------------------
\subsection{Planetary and asteroids ephemerides}
Planetary ephemerides analysis is a very powerful tool to test gravitation and to constrain extensions of GR. They have been used to constrain deviations from GR (see for example \cite{fienga:2008fk,konopliv:2011dq,pitjeva:2005kx,*pitjeva:2013fk,*pitjeva:2013uq,*pitjeva:2014fj,pitjev:2013qv,iorio:2014yu,hees:2014jk,hees:2015sf}). On the other hand, observations of asteroids performed by the GAIA spacecraft~\cite{mouret:2011uq,hees:2015rc} and with radar observations~\cite{margot:2010fk} will also provide nice opportunities to constrain deviations from GR.

No clean tests of dilatonic theory has been performed using planetary or asteroids ephemerides so far. Instead of fitting standard PPN parameters in the equation of motion and in the propagation of light, we think it is more appropriate to fit the parameters summarized by Eqs.~(\ref{eq:param_fond}) and that are related to the fundamental parameters of the theory through this set of equations. Then, from these parameters, one can compute the coefficients defined by Eqs.~(\ref{eq:param_sum}). These will appear in the modified equations of motion given by Eq.~(\ref{eq:EIH_s}). In addition to this, one needs to consider the modification of the Shapiro delay which appear in the ranging observations. The expression of the Shapiro delay is now given by Eq.~(\ref{eq:shapiro}). Finally, there is an additional effect that needs to be taken into account in the reduction of ranging observations: the modifications of the range observables as developed in Sec.~\ref{sec:range}. The range observations used in planetary ephemerides are always two-ways range measurements (DSN station-spacecraft-DSN station). Therefore, it is sufficient to consider the additional term from Eq.~(\ref{eq:range}). This terms depends on the composition of the clocks used in DSN station. Note that astrometric measurements do not need a similar modifications as mentioned in Sec.~\ref{sec:astrometric}. Then, the parameters defined by Eqs.~(\ref{eq:param_fond}) can be fitted with all the observations used with planetary ephemerides. 

Note that in planetary ephemerides analysis, the Sun gravitational parameter ($\tilde G \tilde m$) is a parameter of the fit. On the other hand, the masses of planets appearing in the equations of motion~(\ref{eq:EIH_s}) are fixed to values determined by the motion of satellites (natural or artificial) orbiting around them. Therefore, these masses may also be affected by a violation of the equivalence principle. Indeed, let assume that the mass of the planet $j$ ($\tilde m_j$) is determined through the motion of a satellite $s_j$ orbiting around it. That means that the measured mass is in fact the product $\tilde m_j (1+\delta_j+\delta_{js_{j}})$ which means that if one wants to be completely rigorous, the masses of the planets in Eq.~(\ref{eq:EIH_s}) should be replaced by $\widetilde m_j (1-\delta_j-\delta_{js_{j}})$ where $\widetilde m_j$ is the mass determined by the motion of a satellite $s_j$ around the planets $j$.

%----------------------------------------------
\subsection{Ephemerides of satellites}\label{sec:sat}
  The equations of motion (\ref{eq:EIH_s}) expressed in a barycentric reference frame are adapted to describe the motion of planets and asteroids around the Sun (see previous section) but are not optimal for integrating the motion of satellites around a planet. It is much more suitable to use an expression of the equations of motion centered around the planet. There are two ways of deriving these equations. The first possibility consists in working with barycentric coordinates and work with the planet-satellite vector, whose evolution equation is given by the difference of the barycentric equations of motion of the satellite and of the planet. The other possibility consists in constructing a proper local reference system (with a rescaling of the coordinates) and express the equations of motion in such a local frame. In GR, the theory of reference frames has been developed by Brumberg and Kopejkin~\cite{brumberg:1989fk,brumberg:1991uq,brumberg:1992yj,klioner:1993tn} and by Damour, Soffel and Xu~\cite{damour:1991tp,*damour:1992lp,*damour:1993kc,*damour:1994fd}. This theory of reference frames has been extended to the PPN formalism in \cite{klioner:2000sh,kopeikin:2004qb}. Note that from a theoretical point of view, the use of a local reference frame is much more elegant but in practice, barycentric coordinates are still widely used in practical data analysis. This is the reason why we will present both approaches.

The motion of satellite (natural or artificial) can also be used to constrain modifications of GR. Around the Earth, the motion of the Moon is regularly used to constrain the PPN parameters with Lunar Laser Ranging (LLR) observations~\cite{williams:1996ij,*williams:2004ys,*williams:2009ys,*williams:2012zr,muller:2012ys}. In addition, the LAGEOS and LARES spacecrafts around the Earth are also used to measure relativistic effects such as the Lense-Thirring effect and to constrain GR extensions~\cite{ciufolini:2004uq, *ciufolini:2012kx, *ciufolini:2012sf, *ciufolini:2013rm}. In the same spirit, satellites around other planets (natural and spacecraft) can also be considered to test gravitation. 

In the following, we will develop both approaches: first, we will derive the satellites equation of motion in barycentric coordinates and second, we will make use of a local planetocentric reference frame.

\subsubsection{Barycentric equation of satellites' motion}
The procedure followed to express the equations of motion is similar to the one from~\cite{brumberg:1989fk,brumberg:1992yj}. We decompose each of the terms appearing in Eq.~(\ref{eq:eqm_simpl}) in two parts: one part related to the planet $P$ around which the satellite is orbiting and one part denoted by a bar which denote the external contribution. More precisely, we use
\begin{subequations}\label{eq:pot_div}
 \begin{align}
  W&= U_P +\bar U \, ,\\
  W^i&=  W^i_P + \bar W^i \, , \\
  \Delta &=\Delta_P + \bar \Delta \, ,
 \end{align}
 while we will neglect the $\mathrm{d}\tilde\beta^T$ contributions in this section. As mentioned in Sec.~\ref{sec:eom_pn_s}, this term is important only when $|D''_{i0}|\gg|D'_{i0}|$ and therefore can  be neglected in most situations.
\end{subequations}

The satellite barycentric coordinates with respect to the planet is given by $\bm x=\bm x_S-\bm x_P$ and follow the equations of motion
\begin{align}\label{eq:eqm_bary_s}
% \frac{d^2 x^i}{dt^2}&=-\frac{\tilde G }{r^3}x^i\Bigg[\tilde m_P(1+\delta_S+\delta_{PS})+\tilde m_S(1+\delta_P+\delta_{ST})\Bigg]\nonumber \\
% &\qquad +Q_i+ T^i(\bm x(t),\bm x_S(t))+\delta T^i(\bm x_S,\bm x_P)  \nonumber\\
% &\qquad+\frac{1}{c^2} \sum_{n=1}^3 (\varphi^i_n+ g^i_n)+\tilde g^i_2 \, , 
  \frac{d^2 x^i}{dt^2}&=-\frac{\tilde G }{r^3}x^i\Bigg[\tilde m_P(1+\delta_S+\delta_{PS})+\tilde m_S(1+\delta_P+\delta_{SP})\Bigg]\nonumber \\
 &\qquad +Q^i_P-Q^i_S+ T^i(\bm x(t),\bm x_S(t))+\delta T^i(\bm x_S,\bm x_P)  \nonumber\\
 &\qquad+\frac{1}{c^2} \sum_{n=1}^3 (\varphi^i_n+ g^i_n)+\tilde \varphi^i_2 \, , 
\end{align}
where the first term is the Newtonian force from the planet (corrected because of the violation of the EEP), the second term $Q^i_P-Q^i_S$ represents the correction for the non-geodesic BCRS motion of the planet $P$ and the satellite $S$. Note that for artificial satelites one has $\tilde m_S = 0$. The following term is the standard Newtonian tidal interaction given by
\begin{align}
 T^i(\bm x,\bm x_S)&=\bar U_{,i}(\bm x_S)-\bar U_{,i}(\bm x_P)\\
 &=-\sum_{A\neq P, A \neq S}\tilde G \tilde m_A\left[ \frac{\bm x_{AS} }{r_{AS}^3} - \frac{\bm x_{AP}}{r_{AP}^3}\right]\nonumber \\
 &\approx -\sum_{A\neq P, A \neq S} \frac{\tilde G \tilde m_A}{r_{AP}^3}\left(\bm x -3  \frac{ \bm x_{AP}(\bm x_{AP}.\bm x)}{r_{AP}^2}\right)\nonumber \, ,
\end{align}
with $\bm x_{AS}=\bm x_S-\bm x_A$.  The last term is a correction to the tidal term due to the violation of the EEP. It reads
\begin{align}
 \delta T^i&=-\sum_{A\neq P, A \neq S}\tilde G\tilde m_A \left[ \frac{\bm x_{AS} }{r_{AS}^3}(\delta_S+\delta_{AS}) - \frac{\bm x_{AP}}{r_{AP}^3}(\delta_P+\delta_{AP})\right]\nonumber \\
 &\approx -\sum_{A\neq P, A \neq S}\frac{\tilde G\tilde m_A}{r_{AP}^3} \bm x_{AP} \left(\delta_S+\delta_{AS}-\delta_P-\delta_{AP}\right)\, .
\end{align}
The first relativistic corrections appearing in the equation of motion are due to the gravitational field of the planets and are given by
\begin{subequations}
 \begin{align}
  \varphi^i_1&= \frac{\tilde G \tilde m_P}{r^3}\Bigg[\left(2(\tilde \beta +\tilde \gamma) \frac{\tilde G \tilde m_P}{r} -\tilde \gamma v^2 \right) x^i \nonumber \\
  &\qquad \qquad\qquad+2(1+\tilde\gamma) (\bm x.\bm v) v^i\Bigg] \, ,\\
  g^i_1 &=\frac{\tilde G \tilde m_P}{r^3}\Bigg\{ x^i\left[2\bm v_p .\bm v+v_p^2+\frac{3}{2}\left(\frac{\bm v_p.\bm x}{r}\right)^2\right]+\bm v_P.\bm x~ v^i\Bigg\} \, ,
 \end{align}
 where the first part is the standard PPN Schwarzschild correction and the second term is a term due to the motion of the planet (i.e. due to a Lorentz contraction). We will see that this term is a coordinate effect that can be removed by using a suitable planetocentric reference frame. The second relativistic terms are given by
\begin{align}
 \varphi^i_2&=-2(\tilde\beta+\tilde\gamma)\Bigg\{U_P\left[\bar U_{,i}(\bm x_P+\bm x)-\bar U_{,i}(\bm x_P)\right] \\
 &\quad+ U_{P,i}\left[\bar U(\bm x_P+\bm x)-\bar U(\bm x_P)-\bar U_{,k}(\bm x_P)x^k\right]\Bigg\}\, ,\nonumber \\
 \tilde\varphi_2^i&=-(4\tilde\beta-\tilde\gamma-3)\bar U(\bm x_P) U_{P,i}-\frac{4\tilde\beta-\tilde\gamma-3}{2} U_{P,i} a^k_P x^k \nonumber\\&\qquad -\frac{4\tilde\beta-\tilde\gamma-3}{2}U_P a^i_P \, , \label{eq:tilde_phi2}\\
 g^i_2&=-(3\tilde\gamma+2)\bar U(\bm x_P)U_{P,i} -\frac{5\tilde\gamma+4}{2}U_{P,i}\bm a_P. \bm x-\frac{\tilde \gamma}{2}U_P a^i_P \,,
\end{align}
where $a^i_P=\bar U_{,i}(\bm x_P)-Q^i_P$ is the Newtonian acceleration of the planet. We will neglect terms of the order $Q_i/c^2$.
The first term corresponds to relativistic non-linear couplings which have tidal form. The second term is absent in GR and corresponds to Eq.~(8.7) from \cite{klioner:2000sh} and has already been derived in \cite{shahid-saless:1988sf}. The observability of this term has been studied in~\cite{shahid-saless:1988sf,damour:1994nr}. As we will see, terms in $ g^i_2$ are coordinate effects that can be removed by using a proper local reference frame.

Finally, the last terms are given by
\begin{align}
 \varphi^i_3&=\Bigg[-2(\tilde\beta+\tilde\gamma)\bar U \bar U_{,i}-\bar \Delta_{,i} + \tilde\gamma (v_p^k+v^k)(v_p^k+v^k)\bar U_{,i}\nonumber \\
 &\qquad -2(1+\tilde\gamma)(v_p^k+v^k)\bar W^k_{,i} +2(1+\tilde\gamma) v^k\bar W^i_{,k} \nonumber \\
 &\qquad+ 2 (1+\tilde\gamma)\overset{*}{\bar W^i}-(2\tilde\gamma+1)\overset{*}{\bar U}(v_P^i+v^i) \nonumber \\
 &\qquad-\bar U_{,k}(v_P^i+v^i)(2(1+\tilde\gamma)v^k+v_P^k)\Bigg]_{\bm x_P}^{\bm x_P+\bm x}\, , \\
 g^i_3&=-(1+2\tilde\gamma) v^i \dot{\bar U}(\bm x_P)+\bar U_{,i}(\bm x_P)(2v_P^kv^k+v^2)\nonumber \\
 &-\bar U_{,k}(\bm x_P) \left[2(1+\tilde\gamma) ( v^kv^i+v^i_P v^k)+v^iv^k_P\right] \nonumber\\
 &+2(1+\tilde\gamma)\bar W^i_{,k}(\bm x_P)v^k-2(1+\tilde\gamma)\bar W^k_{,i}(\bm x_P)v^k \, ,
\end{align}
where the dot denots the total time derivative and the star means the total time derivative under constant $x^k$
\begin{equation}
 \left.\dot{\bar U}\right|_{\bm x_P}^{\bm x_P+\bm x}=\left.\overset{*}{\bar U}\right|_{\bm x_P}^{\bm x_P+\bm x}+v^k\bar U_{,k}(\bm x_P+\bm x) \, .
\end{equation}
These equations are a generalization of the ones that can be found in \cite{brumberg:1989fk,brumberg:1992yj}. The detailed expressions of the potentials and their derivatives is given explicitely in Appendix~\ref{app:pot}.
\end{subequations}

%----------------------------------------------------------------------------
\subsubsection{Planetocentric equation of satellites motion}
Instead of working with the equations of motion expressed in terms of barycentric coordinates, one can use local planetocentric coordinates. They have the advantage to be more appropriate to describe the gravitational physics in the surronding of the planet. In the PPN formalism, the coordinate transformation between the BCRS and a local planetocentric reference frame has been derived in \cite{klioner:2000sh,kopeikin:2004qb} and writes
\begin{subequations}\label{eq:trans_coord}
 \begin{align}
  T&=t-\frac{1}{c^2}\left[S_P(t)+v_P^k x^k \right]+\mathcal O(1/c^4) \, , \\
  X^i&= x^i +\frac{1}{c^2}\Bigg[\left(\frac{1}{2} v_P^iv_P^k + qF^{ik}_P+D^{ik}_P\right)x^k \\
  &\hspace{3cm} + D^{ijk}_P x^jx^k\Bigg]+\mathcal O(1/c^4) \nonumber \, ,
 \end{align}
 with\footnote{Note that here, we use the transformation from \cite{klioner:2000sh} which differs from the ones from \cite{kopeikin:2004qb} by the $\tilde\gamma$ coefficient in $D^{ijk}$. This corresponds to a choice of gauge, see the discussion in \cite{kopeikin:2004qb}.}
 \begin{align}
  \dot S_P&=\frac{1}{2}v_P^2 + \bar U(\bm x_P) \, , \\
  D^{ik}_P&=\tilde\gamma \delta_{ik}\bar U(\bm x_P) \, ,\\
  D^{ijk}_P&=\frac{\tilde \gamma}{2}\left(\delta_{ij}a^k_P + \delta_{ik}a^j_P-\delta_{jk}a^i_P \right) \, , \\
  \dot  F^{ik}_P&=\frac{(1+2\tilde\gamma)}{2} \left(v_P^i\bar U_{,k}(\bm x_P)-v_P^k\bar U_{,i}(\bm x_P)\right)\nonumber  \\
  &-(1+\tilde\gamma)\left(\bar W^i_{,k}(\bm x_P)-\bar W^k_{,i}(\bm x_P)\right)\\
  &+\frac{1}{2}v^i_PQ^k_P-\frac{1}{2}v^k_P Q^i_P\nonumber\, .
 \end{align}
 where capital letters design local coordinates. The parameter $q$ defines the type of reference system used: $q=0$ characterizes a kinematically non-rotating system and $q=1$ characterizes a dynamically reference system~\cite{brumberg:1989fk,brumberg:1992yj}.
\end{subequations}

Using this coordinates transformation, one can transform the equations of motion (\ref{eq:eqm_bary_s}) and express them in a local planetocentric reference frame. The procedure used is similar to the one described in Sec.~8 of \cite{brumberg:1989fk} (see also~\cite{brumberg:1992yj}). The first step consists in transforming the left-hand side of the equation of motion by using the transformation of coordinates from Eqs.~(\ref{eq:trans_coord}) to obtain in computing $d^2 X^i/dT^2=d^2 x^i/dt^2 +\Delta^i_1/c^2$. The expression of $\Delta^i_1$ is given in \cite{brumberg:1992yj} (see also Eq. (8.22) of \cite{brumberg:1989fk}). The second step consists in transforming the right-hand sides of Eq.~(\ref{eq:eqm_bary_s}) to the new variables $\bm X$ and $T$ (see Eqs. (8.24-8.29) from \cite{brumberg:1989fk}). This makes appear another correction $\Delta^i_2$ whose expression is given in \cite{brumberg:1989fk,brumberg:1992yj}. Keeping only the leading terms related to the violation of the EEP gives the following planetocentric equations of motion
\begin{align}
 \frac{d^2X^i}{dT^2}&=-\frac{\tilde G }{R^3}X^i\Bigg[\tilde m_P(1+\delta_S+\delta_{PS})+\tilde m_S(1+\delta_P+\delta_{ST})\Bigg]\nonumber \\
 &\qquad+Q^i_P-Q^i_S  +T^i(\bm x_S(t_*),\bm x_P(t_*)) + \delta T^i(\bm x_S,\bm x_P) \nonumber\\
 &\qquad +\frac{1}{c^2}\left(\varphi^i_1+\varphi^i_2 +\tilde\varphi^i_2 + \Phi^i_\textrm{rot}+\Phi^i_\textrm{el}+\Phi^i_\textrm{mg}\right)\, .
\end{align}
The time $t_*$ appearing in the tidal term is the TCB corresponding to the GCRS coordinates $(T,\bm X=0)$, which is solution of $T=t^*-S_P(t^*)/c^2$ (see the discussion in \cite{klioner:1993tn})\footnote{As mentioned in \cite{klioner:1993tn}, one can use $t$ the TCB related to $(T,\bm X)$ in the Newtonian tidal term. Doing this will modify slightly the expression of $\Phi_\textrm{el}$ but leads to complications for practice use.}. A tedious but straightforward calculation (similar to the ones from \cite{brumberg:1989fk,brumberg:1992yj}) allows  one to show that the coordinate transformation compensates exactly the terms $g^i_1$ and $g^i_2$ and to derive the expression of the other terms. The $\Phi^i_\textrm{rot}$ terms are coriolis terms appearing if one express the local equation of motion in a kinematically non-rotating reference frame 
\begin{equation}
	\Phi^i_\textrm{rot}=2(q-1)\dot F_P^{ik} V^k+(q-1)\ddot F_P^{ik} X^k \, . 
\end{equation}
The $\Phi^i_\textrm{mg}$ contains the gravito-magnetic terms that are dependent on the satellite velocity
\begin{widetext}
\begin{subequations}
	\begin{align}
		\Phi^i_\textrm{mg}	&=\mathcal A^i(\bm x_P+\bm X)-\mathcal A^i(\bm x_P)+2\tilde\gamma \dot a^k_P V^k X^i  +\Big[(2\tilde\gamma+1)\dot a^k_P V^i-2\tilde\gamma\dot a^i_PV^k\Big] X^k \, ,
	\end{align}
	where in all the $1/c^2$ terms, the barycentric position $\bm x_P$, velocity $\bm v_P$ and acceleration $\bm a_P$ of the planet  are evaluated at the TCB $t=T$ and with
	\begin{align}
		\mathcal A^i(\bm x)=& \left[2(\tilde\gamma+1)v_p^k+V^k\right] V^k \bar U_{,i}(\bm x)-2(1+\tilde \gamma)v^i_P V^k +2(1+\tilde\gamma)V^k \bar W^i_{,k}-2(1+\tilde\gamma)V^k \bar W^k_{,i} -(1+2\tilde\gamma) \overset{*}{\bar U}(\bm x)V^i \, .
	\end{align}
\end{subequations}	

The last term in the equation of motion $\Phi_\textrm{el}$ is the gravito-electric term that does not depend on the velocity of the satellite. It is given by

\begin{subequations}
	\begin{align}
		\Phi^i_\textrm{el}=&\mathcal B^i(\bm x_P+\bm X)-\mathcal B^i(\bm x) -\Bigg[\frac{1}{2}v_P^kv_P^mX^m+q F_P^{km}X^m + \tilde \gamma \bar U(\bm x_P)X^k +\tilde \gamma X^k X^m a_P^m -\frac{\tilde\gamma}{2}X^m X^m a_P^k \Bigg]\bar U_{,ik}(\bm x_P+\bm X) \nonumber\\
		&\qquad\qquad+\Bigg[a^i_Pa^k_P+(1+\tilde\gamma)\dot{\bar W}^k_{,i}(\bm x_P)-(1+\tilde\gamma)\dot{\bar W}^i_{,k}(\bm x_P)+(1+\tilde\gamma)v^i_P\dot a^k_P-\tilde\gamma v_P^k\dot a^i_P\Bigg]X^k \nonumber\\
		&\qquad\qquad+\tilde\gamma \ddot{\bar U}(\bm x_P)X^i +\tilde\gamma \ddot a_P^k X^k X^i -\frac{\tilde\gamma}{2}\ddot a^i_P X^k X^k +T^i_{,t}(\bm x_P+\bm X,\bm x_P)v_P^k X^k \, , \label{eq:phi_el}
	\end{align}
	with
	\begin{align}
		\mathcal B^i(\bm x)=&-2(\tilde\gamma+\tilde\beta)\bar U(\bm x)\bar U_{,i}(\bm x)-\bar \Delta_{,i}+\Big[(1+\tilde\gamma)v_P^2+(2+\tilde\gamma)\bar U(\bm x_P)+(2+\tilde\gamma)a_P^kX^k\Big]\bar U_{,i}(\bm x) +2(1+\tilde\gamma)\overset{*}{\bar W^i}(\bm x)\nonumber \\
		&+\Bigg[-\frac{1}{2}v^i_Pv^k_P + q F_P^{ik}+\tilde \gamma a^k_PX^i-\tilde\gamma a^i_P X^k\Bigg]\bar U_{,k}(\bm x) -(1+2\tilde\gamma)\overset{*}{\bar U}(\bm x)-2(1+\tilde\gamma)v_P^k\bar W^k_{,i}(\bm x) \, .
	\end{align}
\end{subequations}
\end{widetext}
The expressions of the potentials appearing above and their derivatives are given in Appendix~\ref{app:pot}. The last term in Eq.~(\ref{eq:phi_el}) is due to the choice of the coordinate time $t_*$ at which we evaluate the tidal term in the Newtonian part of the equation of motion (see the discussion in \cite{klioner:1993tn} for further information). In the GR limit, these equations recovers the ones from \cite{brumberg:1989fk,brumberg:1992yj,klioner:1993tn}. Note that the equations presented here are a simplified version of the ones developed in \cite{klioner:2000sh,kopeikin:2004qb} where the full development of a proper reference frame in the PPN formalism taken into account the internal structure of the bodies is performed.

The previous equation can be used to analyse data from observations of satellites (artificial and natural) orbiting around planets. The term given by Eq.~(\ref{eq:tilde_phi2}) vanishes in GR. This term shows that the external gravitational field in which a system is embedded influenced the internal dynamics of the systems (in addition to the standard tidal interaction). This feature is a characteristic of a violation of the strong equivalence principle. In our case, it appears as a coupling between the satellite orbit and the orbit of the planet around which it orbits. This term allows one to constrain gravitation as mentioned in \cite{damour:1994fd}. Nevertheless, one has to be careful since the leading term in $\varphi_2^i$ is highly correlated to the mass of the planet. Let us illustrate this fact by taking a planet $P$ orbiting around a central mass $\tilde M$ with $\bm D$ the position of the planet with respect to this central mass. Then, the first term from Eq.~(\ref{eq:tilde_phi2}) is given by
\begin{equation}\label{eq:main_Add}
 (4\tilde\beta-\tilde\gamma-3) \frac{\tilde G \tilde M }{Dc^2} \frac{\tilde G \tilde m_P}{r^3}x^i \, .
\end{equation}
In the case where the planet $P$ has a nearly circular orbit around the central mass, this term can be absorbed in a redefinition of the planet mass $\tilde m_P$ and therefore is difficult to measure. It is therefore highly important to fit the mass of the planet in this kind of analysis and to be careful about these expected correlations.

Note that the equations of motion are only one part of the modifications that need to be included in the data analysis. One needs also to  consider the effects on the propagation of light as mentioned in Sec.~\ref{sec:light} and the modification of the expression of the observables, especially in the case of a range measurement (see Sec.~\ref{sec:range}), which is common for Lunar Laser Ranging and ranging to spacecraft (around Earth and around other planets).

Finally, since the external gravitational field in which a system is embedded can have an impact on its internal dynamics, it is fair to wonder if one can use the effect of our Galaxy on the Solar System dynamics. In the PPN formalism, our Galaxy will produce an additional term in the planetary equation of motion given by Eq.~(\ref{eq:tilde_phi2}). The leading term is highlighted in Eq.~(\ref{eq:main_Add}). Unfortunately, even if the order of magnitude of this term is above the current observations accuracy, it is completely correlated to the Sun mass and is not measurable. The other terms in Eq.~(\ref{eq:tilde_phi2}) are smaller by a ratio $r/D$ and therefore are too small to constrain the PPN parameters.

\subsection{Very Long Baseline Interferometry}
Very Long Baseline Interferometry (VLBI) observations can also be used to constrain alternative gravitation theories~\cite{lambert:2009bh,*lambert:2011ap}. The observable quantities in VLBI observations are recorded signals measured in the proper time of station clocks. A correlator then transform signals into a difference of coordinate Terrestrial Times (TT) related to two reception events of the same signal. If we denote by 1 and 2 the two stations that record an electromagnetic signal, the observables are $\hat \tau_1$ and $\hat \tau_2$ given by the integration of Eq.~(\ref{eq:dhattau_dt_simp}). In the standard procedure described into detail in Chapter 11 of the International Earth Reference System conventions~\cite{petit:2010fk}, these proper times are transformed into TT times and afterwards into Barycentric Coordinate Times (TCB). The difference of the two TCB times can be computed from the incident direction of the electromagnetic signal and include the differential Shapiro delay~\cite{petit:2010fk}.

The dilaton theories considered in this paper will produce two effects that may impact VLBI observations. First, the propagation of light will be modified as mentioned in Sec.~\ref{sec:light}, which will produce a modifiaction of the differential Shapiro delay that can be parametrized by the standard PPN parameter $\tilde\gamma$ (and if one wants to be rigorous by a rescaling of the mass, see the discussion in Sec.~\ref{sec:light}). In addition to this standard effect, the violation of the EEP will affect the time measured by real clocks that is now given by Eq.~(\ref{eq:dhattau_dt_simp}). Therefore, the transformation between the time given by the two local clocks and TCB times will now be given by the solution of 
\begin{equation}\label{eq:transf_time}
 \hat \tau_i= \tau_\textrm{PPN} (t_i) + \delta\hat\tau_i(t_i) \, ,
\end{equation}
where $\tau_\textrm{PPN}$ is the expression of the proper time as a function of TCB in the standard PPN formalism  (note that the $\tilde\gamma$ parameter enters the coordinate transformation between the geocentric reference frame and the barycentric one~\cite{petit:2010fk} as can be seen from Eqs.~(\ref{eq:trans_coord})). The second term of the last expression is the contribution coming from the violation of the EEP which is given by the integration of
\begin{equation}
 \frac{d\delta\hat\tau_i}{dt}= -\sum_A \frac{\tilde G \tilde m_A}{c^2 r_{Ai}} \left(\hat \delta_i + \hat\delta_{iA} \right) \, ,
\end{equation}
where the expressions of the $\hat \delta$ coefficients are given by Eqs.~(\ref{eq:hat_deltas}). If one makes a decomposition of the TCB as $t_i=t_{i,\textrm{PPN}}+\delta \hat t_i$ where $t_{i,\textrm{PPN}}$ is the TCB computed in the standard PPN formalism solution of $\hat \tau_i= \tau_\textrm{PPN} (t_{i,\textrm{PPN}})$ and the $\delta \hat t_i$ terms is coming from the violation of the EEP. An expansion of Eq.~(\ref{eq:transf_time}) leads to 
\begin{equation}
 \delta \hat t_i=-\frac{\delta \hat\tau_i(t_{i,\textrm{PPN}})}{\left.\frac{d\tau_\textrm{PPN}}{dt}\right|_{t_{i,\textrm{PPN}}}}\approx - \delta \hat\tau_i(t_{i,\textrm{PPN}})+\mathcal O(1/c^2)\, .
\end{equation}
Therefore, the modeling of the TCB time difference in the dilaton theories is now given by
\begin{equation}
 t_1-t_2=\left[t_1-t_2\right]_\textrm{PPN} + \delta \hat \tau_2  - \delta \hat\tau_1 \, ,
\end{equation}
where the first part is the standard PPN model~\cite{petit:2010fk} that includes the $\tilde\gamma$ contribution in the transformation between reference frames  and the Shapiro delay parametrized by the same PPN parameter. The second part modeled the violation of the EEP and depends on the composition of the two clocks. The amplitude of this second modification depends on how and how often VLBI local clocks are synchronized.

   %-------------------------------------------------------------------------------------
\section{Specific cases}\label{sec:specific}
In this section, we will take some particular cases of the action (\ref{eq:actionstringframe}) and show we recover standard results. Moreover, the case of the decoupling limit presented in Sec.~\ref{sec:decoupling} will be considered into detail.

  %-------------------------------------------------------------------------------------
\subsection{Brans-Dicke-like theories}
\label{sec:BD}
Brans-Dicke-like theories are characterized by $f(\varphi)=\varphi$, $\omega(\varphi)$ and all $D_i(\varphi)=0$ (no violation of the Einstein equivalence principle) which leads to $Z(\varphi)=(2\omega(\varphi)+3)/2\varphi^2$. This leads to 
\begin{equation}
 \tilde \gamma=\frac{1+\omega_0}{2+\omega_0} \, ,
\end{equation}
which the standard result~\cite{brans:1961fk,brans:2014sc}. Similarly the observed Newton constant is given by
\begin{equation}
\tilde G=\frac{G}{\varphi_0}\left(1+ \frac{1}{2Z_0}\left(\frac{f'_0}{f_0}\right)^2\right)=\frac{G}{\varphi_0}\frac{2\omega+4}{2\omega+3} \, ,
\end{equation}
which is standard too. And finally, the PPN $\beta$ is given by
\begin{eqnarray}
 \tilde \beta
 &=&1 + \frac{\varphi_0 \omega'_0 }{4\left(\omega_0+2\right)^2\left(2\omega_0+3\right)} \, .
\end{eqnarray}
 %-------------------------------------------------------------------------------------
\subsection{Damour and Donoghue}
\label{sec:DDcase}
The Damour-Donoghue action is characterized by $f(\varphi)=1$, $\omega(\varphi)=2\varphi$ which leads to $Z(\varphi)=2$ and the masses depends linearly on the scalar field so that $\tilde \beta_{A0}=0$. The expressions from \cite{damour:2010zr,damour:2010ve} are recovered from the formulas developed in this paper. In particular, one finds to the expression of the observed gravitational constant given by
\begin{equation}
	\tilde G=G(1+{d^*_g}^2) \, ,
\end{equation}
and a modification of the Newtonian potential characterized by
\begin{equation}
	\tilde W_T =\sum_A \frac{\tilde G\tilde m_A}{r_{AT}}\left(1+d_g^*\bar \alpha_T+\bar \alpha_T\bar \alpha_A \right) \, .
\end{equation}
The relative acceleration between two test particles in an external gravitational field generated by a body $E$ therefore writes
\begin{equation}
\left(\frac{\Delta a}{a} \right)_{BC} \approx (d^*_g+\bar{\alpha}_{E0}) (\bar{\alpha}_{B0}-\bar{\alpha}_{C0}) \, .
\end{equation}
At the same time, the PPN parameter is given by
\begin{equation}
 \tilde \gamma = \frac{1 - {d^*_g}^2}{1 + {d^*_g}^2} \, .
\end{equation}
Therefore, from observational constraints on the post-Newtonian $\tilde\gamma$ parameter ($|1-\tilde\gamma|\lesssim 10^{-5}$ \cite{bertotti:2003uq}) one knows that $d^*_g \lesssim 2.10^{-3}$.

%-------------------------------------------------------------------------------------
\subsection{Phenomenology of a model with a decoupling}
\label{sec:Phendecoupling}
As already mentioned in Sec.~\ref{sec:decoupling}, a decoupling scenario occurs when the scalar-Ricci coupling (the function $f(\varphi)$ in the action) and the scalar-matter coupling (characterized by the functions $D_i(\varphi)$) are related through Eq.~(\ref{eq:decoupling}). This decoupling is similar to the decoupling studied in \cite{minazzoli:2013fk} where a simple effective multiplicative coupling between the scalar field and the rest mass energy density was considered.

In the framework of dilaton theories, the decoupling introduced in Sec.~\ref{sec:decoupling} leads to $\alpha_0=\beta_0=0$, which means that the PPN parameters take exactly their GR values $\tilde\gamma=\tilde\beta=1$. This result is independent of the function $\omega(\varphi)$ appearing in the action~(\ref{eq:actionstringframe}). In addition, the parameters encoding the leading terms related to a violation of the UFF $\delta_A$ vanishes exactly as well as can be seen from Eq.~(\ref{eq:delta_A}). In this case, the violation of the UFF will be characterized by $\delta_{AT}$ as can be noticed from the Newtonian part of Eq.~(\ref{eq:EIH_s}). Therefore, the $\eta$ parameter introduced in Sec.~\ref{sec:UFF} to characterize a violation of the UFF between two bodies $B$ and $C$ falling in a gravitational field generated by $E$  becomes
\begin{equation}\label{eq:eta_dec}
  \eta_\textrm{dec} \approx \left|\delta_{EB}-\delta_{EC} \right|= \left|\tilde \alpha_{E0}(\tilde\alpha_{B0}-\tilde\alpha_{C0}) \right| \, ,
\end{equation}
where the coefficients $\tilde \alpha_{A0}$ are defined by Eq.~(\ref{eq:tilde_alpha}). In this particular case, a violation of the UFF is still expected even though the values of the PPN parameters are exactly the same as in GR. 

This decoupling and several variations of it are studied in details in \citep{minazzoli:2016rm}.

 %-------------------------------------------------------------------------------------
\section{Conclusion}\label{sec:conclusion}

In this paper, we have derived the post-Newtonian phenomenology of a general dilaton theory where the dilaton-matter coupling is the microscopic model proposed by Damour and Donoghue~\cite{damour:2008pr,damour:2010zr,*damour:2010ve}. In addition, we consider a general dilaton-Ricci coupling and a general dilaton kinetic term. This generalization allows one to exibit a decoupling scenario studied in details in \cite{minazzoli:2016rm}. This scenario turns out to be interesting in the sense that observational deviations from GR predictions are strongly reduced. In particular, the standard PPN parameters take exactly their GR values and violations of the UFF is strongly reduced. Therefore it may play a role in the explanation of the non detection of a UFF violation so far; while a lot of theoretical developments seem to point out that this principle should be violated at some level~\cite{damour:2012zr}.

In addition, we presented all the equations needed in order to analyze observations within a dilaton framework in order to produce constraints on this GR extension. As mentioned in the introduction, dilaton theories produce two types of effects: violations of the EEP and modifications parametrized by the PPN formalism. Both these effects need to be taken into account simultaneously in the data reduction in order to produce clean constraints on this type of theory. We therefore presented equations of motion of celestial bodies in a barycentrice reference frame (useful for planetary and asteroids ephemerides) and in a planetocentric reference frame (useful for the motion of satellites -- artificial or natural -- around a planet), the equations of the light propagation which has implication on light deflection, on range and Doppler and finally the equation of evolution of the proper time delivered by a specific clock. All these equations can be used in order to analyze data like for example: clocks comparison in order to test the gravitational redshift, planetary ephemerides analysis, development of asteroids ephemerides, motion of artificial satellites around Earth, Lunar Laser Ranging, spacecraft tracking, Very Long Basline Interferometry, tests of the universality of free fall \dots

We insist on the fact that different types of experiment are sensitive to different fundamental parameters. For example, tests of the UFF are highly sensitive to the coupling between the dilaton and the quark masses (characterized by the parameters $d_{m_u}$ and $d_{m_d}$) and to the QCD mass scale (characterized by the parameter $d_g$); while experiments using clocks are more sensitive to the coupling with the electron mass ($d_{m_e}$) and coupling with the fine structure constant ($d_e$). It is therefore important to keep doing different types of experiment in order to derive the stringent constraints on this kind of theory.

It has to be mentioned however that in this study we used the usual dust field approximation in order to model matter. This is motivated by the work of Damour and Donoghe~\cite{damour:2010zr} which derived the link between the the (semi-)fundamental matter Lagrangian (\ref{eq:Lint}) and its effective point particles realization (\ref{eq:smat}).  But because of the non-minimal scalar-matter coupling, one can expect non-trivial effects when considering fluids with pressure \cite{minazzoli:2013fk,minazzoli:2015ax,minazzoli:2014ao,*minazzoli:2014pb}. A careful extension of the results presented in this study for pressureful cases could be the next step in the study of dilaton post-Newtonian phenomenology. However several theoretical issues have to be addressed first. Indeed, because of the non-minimal scalar-matter coupling, parts of the material Lagrangian appear explicitly in the field equations and it is therefore a difficult task to derive from first principle the value of these on-shell parts when one eventually wants to consider effective pressureful fluids. This problem is left for further studies.

%-------------------------------------
\begin{acknowledgments}
The authors are thankful to P. Wolf for his helpful comments on a preliminary version of this manuscript and to A. Fienga for motivating this work.

\end{acknowledgments}

\bibliography{../biblio_COPY}

\appendix

%------------------------------------------
\section{More accurate coupling coeficients}\label{app:alphacoeff}

Equation (\ref{eq:alpha_DD}) is an approximation that is suitable most of the time. However, it is not suitable for nucleons or for light isotopes for instance. Indeed, $Z/A = 1/2$ has been used to derive (\ref{eq:alpha_DD}). Therefore, it is important to remind what the most complete formulae is. When the contributions of strange quarks are neglected it is given in \cite{damour:2010zr,*damour:2010ve}. If one implements non-linear couplings as well, it becomes:
\begin{subequations} \label{eq:alpha_DD_full}
\begin{equation}
\label{alpha5}
{\alpha}_A = D_g'(\varphi) +  \bar{\bar{\alpha}}_A
\end{equation}
with the decomposition
\begin{eqnarray}
\bar{\bar{\alpha}}_A =  [&& (D_{\hat m}'(\varphi) - D_g'(\varphi)) Q_{\hat m} + (D_{\delta m}'(\varphi) -D_g'(\varphi)) Q_{\delta m} \nonumber \\
&&+ (D_{m_e}'(\varphi) - D_g'(\varphi)) Q_{m_e} + D_e'(\varphi) Q_e ]_A, \label{baralpha4}
\end{eqnarray}
where $Q_{k_a}$ are given by
\begin{eqnarray}
 Q_{\hat m} = F_A \Big[&& 0.093 -\frac{0.036}{A^{1/3}} - 0.020 \frac{(A-2Z)^2}{A^2} \nonumber\\
&& - 1.4 \times 10^{-4} \, \frac{Z(Z-1)}{A^{4/3}} \Big]  ,\label{Qmhat}
\end{eqnarray}
\begin{equation}
\label{Qdeltam}
Q_{\delta m} = F_A \left[0.0017   \, \frac{A-2Z}{A} \right] ,
\end{equation}
\begin{equation}
\label{Qme}
 Q_{m_e} = F_A \left[ 5.5 \times 10^{-4}  \, \frac{Z}{A} \right] ,
\end{equation}
and
\begin{equation}
\label{Qe}
Q_e = F_A   \left[  -1.4 + 8.2 \frac{Z}{A} + 7.7 \frac{Z(Z-1)}{A^{4/3}}  \right]\times 10^{-4}.
\end{equation}
The factor $F_A$ denotes $F_A \equiv A m_{\rm amu}/m_A$ (it can be replaced by one in lowest approximation) \cite{damour:2010zr,*damour:2010ve}. One recovers Eqs. (\ref{eq:alpha_NL}) by setting $Z/A=1/2$, neglecting the third term in (\ref{Qmhat}), neglecting (\ref{Qdeltam}) and assuming $D_{\delta m}'(\varphi) \sim D_{\hat m}'(\varphi) \sim D_{m_e}'(\varphi)$.
\end{subequations}
Since in most of the paper we parametrize the coupling coefficients by $D'^*_g(\varphi)$ and $\bar \alpha_P$ (Eqs. (\ref{eq:alpha_NL})), it is interesting to give the link between the two parametrizations. It simply reads
\begin{eqnarray}
\bar \alpha_A&=& \bar{\bar{\alpha}}_A+ D'_g(\varphi)-D'^*_g(\varphi) \\
&=&\bar{\bar{\alpha}}_A - 0.093\big(D_{\hat{m}}'(\varphi)-D_g'(\varphi)\big)-0.000 \ 27 D_e'(\varphi). \nonumber
\end{eqnarray}

%------------------------------------------
\section{Quantities appearing in the equations of satellite motion}\label{app:pot}
The goal of this section is to write explicitly all the quantities appearing in the satellite equations of motion derived in Sec.~\ref{sec:sat}. First of all, the decomposition of the potentials from Eq.~(\ref{eq:pot_div}) leads to the following expressions
\begin{subequations}
	\begin{align}
		U_P(\bm x)=&\frac{\tilde G\tilde m_P}{r_P} \, ,\\
		\bar U(\bm x)=&\sum_{A\neq P}\frac{\tilde G\tilde m_A}{r_A}\, ,\\
		W^i_P(\bm x)=&\frac{\tilde G\tilde m_P}{r_P}v^i_P \, , \\
		\bar W^i(\bm x)=&\sum_{A\neq P}\frac{\tilde G\tilde m_A}{r_A}v^i_A \, \\
		\Delta_P(\bm x)=&\frac{\tilde G\tilde m_P}{r_P}\Bigg[-(1+\tilde\gamma)v_P^2+\frac{1}{2}\bm r_P.\bm a_P\nonumber \\
		& +\frac{1}{2}\left(\frac{\bm r_P.\bm v_P}{r_P}\right)^2 +(2\tilde\beta-1)\bar U(\bm x_P)\Bigg] \, ,\\
		\bar \Delta(\bm x)=&\sum_{A\neq P}\frac{\tilde G\tilde m_A}{r_A}\Bigg[-(1+\tilde\gamma)v_A^2+\frac{1}{2}\bm r_A.\bm a_A\nonumber \\
		& +\frac{1}{2}\left(\frac{\bm r_A.\bm v_A}{r_A}\right)^2 +(2\tilde\beta-1)\bar U(\bm x_A)\Bigg] \, ,
	\end{align}
	where  $\bm r_P=\bm x-\bm x_P$ and $r_P=\left|\bm x-\bm x_P\right|$.
\end{subequations}

The following derivatives appear in the equations of motion
\begin{subequations}
	\begin{align}
		U_{P,i}(\bm x)=&-\frac{\tilde G \tilde m_P}{r_P^3}r_P^i \, , \\
		U_{P,t}(\bm x)=&\frac{\tilde G \tilde m_P}{r_P^3}r_P^k v^k_P \, , \\
		\dot U_P(\bm x)=&-\frac{\tilde G \tilde m_P}{r_P^3}r_P^k (v^k-v_P^k) \, .
	\end{align}
	The derivatives of the vector potential are given by
	\begin{align}
		W^i_{P,k}(\bm x)=&-\frac{\tilde G \tilde m_P}{r_P^3}r_P^kv_P^k \, \\
		\dot W^i_P(\bm x)=&-\frac{\tilde G\tilde m_P}{r_P^3}r_P^k (v^k-v_P^k) v^i_P+\frac{\tilde G\tilde m_P}{r_P}a^i_P \, ,
	\end{align}
	while the derivative of $\Delta_P$ is given by
	\begin{align}
		\Delta_{P,i}(\bm x)=&\frac{\tilde G\tilde m_P}{r_P^3}r_P^i\Bigg[(1+\tilde\gamma)v_P^2-\frac{1}{2}\bm r_P.\bm a_P-(2\tilde\beta-1)\bar U(\bm x_P)\nonumber\\
		&+\frac{3}{2}\left(\frac{\bm r_P.\bm v_P}{r}\right)^2\Bigg]+\frac{\tilde G\tilde m_P}{r^3}r_P^kv_P^kv_P^i+\frac{\tilde G\tilde m_P}{2r_P}a^i_P \, .
	\end{align}
\end{subequations}
The derivatives of the bare potentials are similar
\begin{subequations}
	\begin{align}
		\bar U_{,i}(\bm x)=&-\sum_{A\neq P}\frac{\tilde G\tilde m_A}{r^3_A}(x^i-x^i_A)\, , \\
		\dot {\bar U}(\bm x) =&-\sum_{A\neq P}\frac{\tilde G \tilde m_A}{r_A^3}r_A^k (v^k-v_A^k) \, , \\
		\ddot {\bar U}(\bm x)=&-\sum_{A\neq P}\frac{\tilde G \tilde m_A}{r_A^3}\Bigg[ (v^k-v_A^k)(v^k-v_A^k)\nonumber \\
		&+r_A^k(a^k-a_A^k)-3\left(\frac{r_A^k(v^k-v_A^k)}{r_A}\right)^2\Bigg] \, ,
	\end{align}
	\begin{align}
		\bar W^i_{,k}(\bm x)=&-\sum_{A\neq P}\frac{\tilde G \tilde m_A}{r_A^3}r_A^k v_A^i \, , \\
		\dot {\bar W}^i(\bm x)=&-\sum_{A\neq P}\frac{\tilde G\tilde m_A}{r_A^3}r_A^k (v^k-v_A^k) v^i_A+\frac{\tilde G\tilde m_A}{r_A}a^i_A \, , \\
		 \dot{\bar W}^i_{,k}=& -\sum_{A\neq P}\frac{\tilde G \tilde m_A}{r_A^3}\left[r_A^k a_A^k + (v^k-v_A^k)v_A^k\right]  \\
		 &\quad +3 \sum_{A\neq P} \frac{\tilde G \tilde m_A}{r_A^5}r_A^k v_A^i r_A^j (v^j-v_A^j)   \, ,
	\end{align}
	and
	\begin{align}
		\bar \Delta_{,i}(\bm x)=&\sum_{A\neq P}\frac{\tilde G\tilde m_A}{r_A^3}r_A^i\Bigg[(1+\tilde\gamma)v_A^2-(2\tilde\beta-1)\bar U(\bm x_A)\nonumber\\
		&\qquad -\frac{1}{2}\bm r_A.\bm a_A+\frac{3}{2}\left(\frac{\bm r_A.\bm v_A}{r_A}\right)^2\Bigg] \nonumber \\
		&+\sum_{A\neq P}\frac{\tilde G\tilde m_A}{r_A^3}r_A^k v_A^k v_A^i+\sum_{A\neq P}\frac{\tilde G\tilde m_A}{2r_A}a^i_A \, .
	\end{align}
\end{subequations}
The expression of the acceleration of the planets $P$ and their time derivatives are also required
\begin{subequations}
 \begin{align}
   a^i_P=&\bar U_{,i}(\bm x_P) -Q^i = -\sum_{A\neq P}\frac{\tilde G\tilde m_A}{r^3_{AP}}r^i_{AP} -Q^i \, , \\
   \dot a^i_P=& -\sum_{A\neq P}\frac{\tilde G\tilde m_A}{r^3_{AP}}\left[ v^i_{AP} - 3 \frac{r_{AP}^i r_{AP}^kv_{AP}^k}{r_{AP}^2} \right]-\dot Q^i \, , \\
    \ddot a^i_P=& -\sum_{A\neq P}\frac{\tilde G\tilde m_A}{r^3_{AP}}\Bigg[ a^i_{AP} - 6 \frac{v_{AP}^i r_{AP}^kv_{AP}^k}{r_{AP}^2}- 3 \frac{r_{AP}^i v_{AP}^2}{r_{AP}^2}  \nonumber \\
    & -3r^i_{AP}\frac{r_{AP}^ka^k_{AP}}{r^2_{AP}}+15r^i_{AP}\left(\frac{r_{AP}^kv^k_{AP}}{r_{AP}^2}\right)^2\Bigg] -\ddot Q^i \, .
 \end{align}
\end{subequations}

\end{document}